\documentclass[aps,prd,tightenlines,preprintnumbers,showpacs,superscriptaddress,nofootinbib,eqsecnum,floatfix,amsmath,longbibliography,
twocolumn]{revtex4-2}

\usepackage{amssymb}

\usepackage{bm}
\usepackage{soul}
\usepackage{graphicx}
\usepackage[pdftex]{color}
\usepackage[dvipsnames]{xcolor}

\definecolor{dblue}{rgb}{0.0, 0.0, 0.6}
\definecolor{ddblue}{rgb}{0.0, 0.0, 0.4}
\definecolor{dgreen}{rgb}{0.0, 0.4, 0.0}
\definecolor{dgray}{gray}{0.4}
\definecolor{ddgray}{gray}{0.3}

\usepackage{hyperref}

\def\floatcaption#1#2{ \caption{#2 \label{#1}}}

\def\cbo{{\,\raise-.15ex\Sc [\,}}                       % curly "
\def\svev#1{\left\langle #1\right\rangle}       % variable < >
\def\ddt#1{{\buildrel {\hbox{\LARGE .\kern-2pt.}} \over {#1}}}% double dot-over
\long \def \blockcomment #1\endcomment{}
\def\be{\begin{equation}}
\def\ee{\end{equation}}
\def\bea{\begin{eqnarray}}
\def\eea{\end{eqnarray}}

\def\gGF{g^2_{\text{GF}}}

\def\svev#1{\left\langle #1\right\rangle}

\begin{document}

\title{Emergent strongly coupled ultraviolet fixed point in four dimensions with 8 K\"ahler-Dirac fermions}
\author{Anna Hasenfratz}
\email{Anna.Hasenfratz@colorado.edu}
\affiliation{Department of Physics, University of Colorado, Boulder, CO 80309, USA}

%\date{\today}

\begin{abstract} % complete
The  existence of a strongly coupled ultraviolet fixed point in 4-dimensional lattice models as they cross into the conformal window has long been hypothesized.  The   SU(3) gauge system  with 8 fundamental fermions is a good candidate to study this phenomenon as it  is expected to be very close to the opening of the conformal window. I study the system  using staggered lattice fermions in the chiral limit. My numerical simulations employ improved  lattice actions that include heavy Pauli-Villars (PV) type bosons. This modification does not affect the infrared dynamics but greatly reduces the ultraviolet fluctuations,  thus allowing the study of stronger renormalized couplings than previously possible. I consider two different PV actions and find that  both  show an apparent continuous phase  transition in the 8-flavor system. 

I investigate the critical behavior  using finite size scaling of the renormalized gradient flow coupling. The finite size scaling  curve-collapse analysis predicts a first order phase transition consistent with discontinuity exponent $\nu=1/4$ in the system without PV bosons. The scaling analysis with the PV boson actions is not consistent with  a first order phase transition.  The numerical data are well described  by  "walking scaling"  corresponding to a renormalization group $\beta$ function that just touches zero, $\beta(g^2) \sim (g^2 - g^2_\star)^2$, though second order scaling cannot be excluded.   Walking scaling could imply that the 8-flavor system is the opening of the conformal window, an exciting possibility that could be related to t'Hooft anomaly cancellation of the system.

\end{abstract}

\maketitle
%%%%%%%%%%%%%%%%%%%%%%%%%%%%%%%%%%%%%%%%%%%%%%%%%%%%%%%%%%%%%%

%%%%%%%%%%%%%%%%%%%%%%%%%%%%%%%%%%%%%
\newpage
\section{\label{intro} Introduction} % Complete
%%%%%%%%%%%%%%%%%%%%%%%%%%%%%%%%%%%%%

The existence of a conformal phase  in 
4-dimensional  gauge-fermion systems  
is well established \cite{Banks:1981nn}.
 The upper limit of the fermion number where the conformal phase becomes infrared free can be predicted perturbatively. However, the  lower limit   where the  QCD-like chirally broken and confining phase turns conformal requires  a non-perturbative approach. 
 The conformal phase is characterized by an infrared fixed point (IRFP) where the gauge coupling is irrelevant.  In the ``walking technicolor" scenario the  conformal window opens up when the  renormalization group (RG) $\beta$ function vanishes just before chiral symmetry breaking could decouple the fermions \cite{Miransky:1998dh,Appelquist:1997fp,Appelquist:2010gy,Golterman:2016lsd,Golterman:2018mfm}. This fixed point turns into an IRFP as the number of fermions is further increased, but at the sill the $\beta$ function has a maximum that just touches zero.     Alternatively, the IRFP could appear together with an ultraviolet fixed point (UVFP)  as hypothesized by several authors \cite{Kaplan:2009kr,Vecchi:2010jz,Gorbenko:2018ncu,Pomarol:2019aae}.  At the sill of the conformal window the $\beta$ function is similar to that of the walking scenario. The difference becomes evident in the conformal regime where the $\beta$ function exhibits both an IRFP and a UVFP. Conformality is "lost" as the two fixed points merge and move into the complex plane. This latter scenario is particularly interesting as the emergence of a  UVFP implies a second order phase transition with a new relevant operator.   The corresponding phase diagram   of  the conformal regime  was sketched in Ref.\cite{Hasenfratz:2019puu} and reproduced in the left panel of Fig.\ref{fig:phasediag}. The sketch shows the  phase structure and the RG flows in an extended parameter space that includes a new relevant operator denoted by $G$. As the number of flavors increases the IRFP moves to weaker coupling. When the Gaussian FP (GFP) and the IRFP merge, the system becomes infrared free. On the other hand, when the number of flavors decreases, the IRFP approaches the UVFP.  The opening of the conformal window is characterized by the merge of the UV and IR fixed points, as is sketched on the right panel of Fig.\ref{fig:phasediag}. 

%%%%%%%%%%%%%%%%%%%%%%%%%%%%%%%%%%%%%%%%%%%%%%%
\begin{figure*}[tbh]
\vspace{-19mm}
%\hspace{-15mm}
\includegraphics[width=0.99\columnwidth]{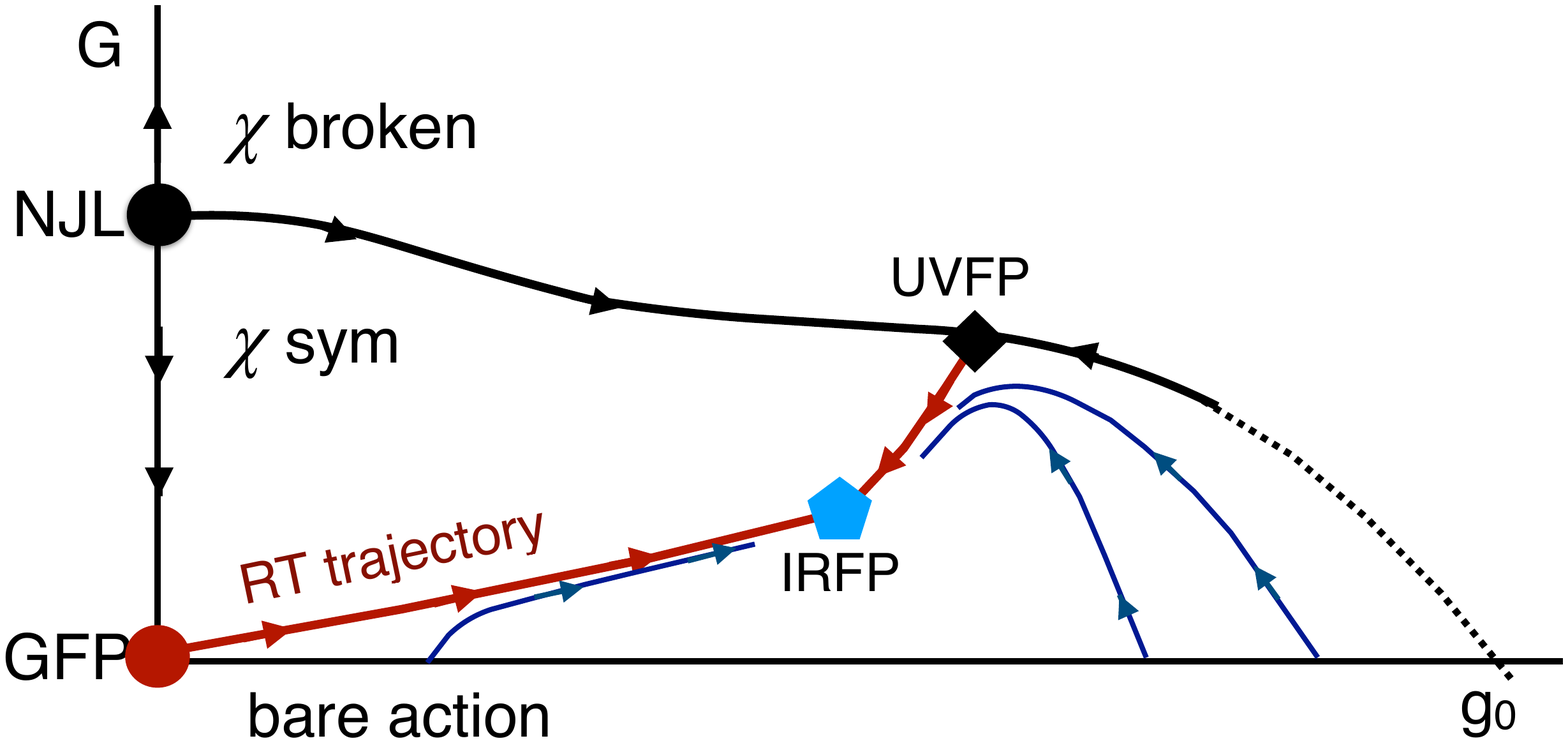} 
\includegraphics[width=0.99\columnwidth]{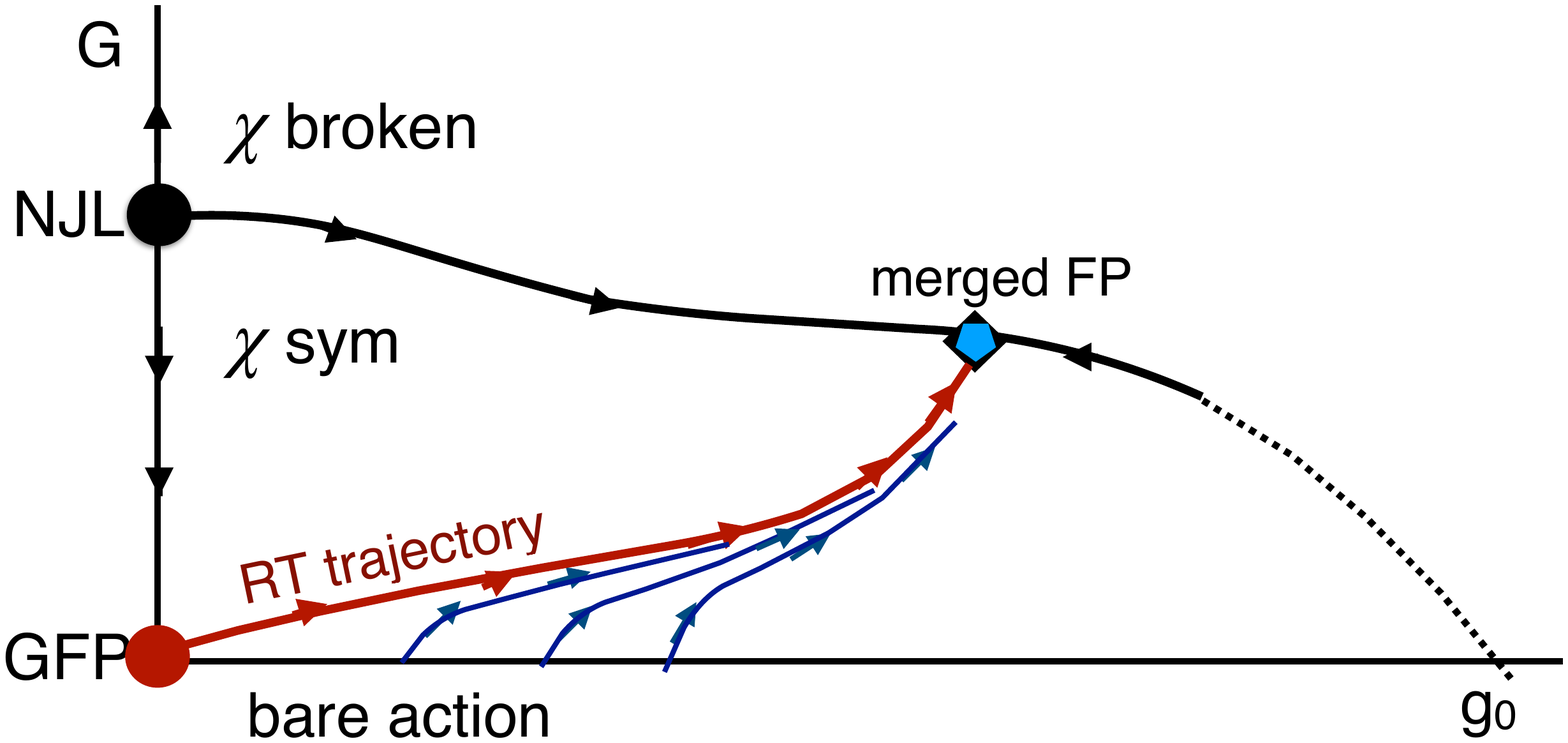} 
\vspace{-7mm}
\floatcaption{fig:phasediag}
{
Sketch of a hypothetical phase diagram  of a conformal system with concomitant infrared and ultraviolet fixed points in an extended parameter space.
$g_0$ refers to the bare gauge coupling, while $G$ denotes the coupling of the emergent relevant operator (possibly a 4-fermion Nambu-Jona-Lasinio interaction).  The left panel, showing separate IR and UV fixed points in a conformal system, is reproduced from Ref. \cite{Hasenfratz:2019puu}.
The right panel shows the "merging" of the two fixed points and corresponds to the sill of the conformal window. The relevant fermion mass  is assumed to be zero, i.e. the sketches describe systems on the critical chiral surface.}
 \end{figure*}
%%%%%%%%%%%%%%%%%%%%%%%%%%%%%%%%%%%%%%%%%%%%%%%
 
The conformal IRFP does not separate phases. However,  a UVFP  would be associated with a continuous phase transition,  indicated by the solid black curves in Fig.\ref{fig:phasediag}.  The phase transition might turn to first order before it intercepts the $G=0$  action, if the two lines intercept at all (dashed black curves in Fig.\ref{fig:phasediag}.).  Phase transitions are much easier to identify in numerical simulations than RG fixed points. QCD-like, chirally broken lattice models  often do not show any phase transition at zero temperature. When they do (e.g. with Wilson fermions), it is a bulk first order transition that arises from lattice artifacts. Conformal lattice systems, on the other hand, almost always exhibit a phase transition from the conformal phase at weak coupling  to a confining one in the strong coupling regime. 
However, despite  extensive numerical studies of systems with various fermion flavors, representations, and gauge colors, no continuous  phase transition has been identified in four dimensional gauge-fermion systems at zero temperature. For a recent review see e.g. \cite{Nogradi:2016qek} and references therein.

In this paper, I present  numerical results that indicate a continuous phase transition of an SU(3) gauge model with 8 fundamental massless fermions 
 implemented as two sets of  staggered lattice fields.
This system has been investigated by several collaborations using different versions of staggered fermions~\cite{Deuzeman:2008sc,Deuzeman:2009mh,Jin:2009mc,Fodor:2009wk,Hasenfratz:2013uha,Aoki:2013xza,Aoki:2014xpa,Hasenfratz:2014rna,Schaich:2015psa,Fodor:2015baa,LatticeStrongDynamics:2018hun,inprepNf8:2022}. Lattice results indicate that $N_f=8$  is  close to the conformal sill, though  there is no solid evidence that would show if it is  inside or out  of the conformal window. For example,  finite temperature simulations at zero fermion mass do not identify a chirally broken phase before the appearance of a first order bulk transition \cite{Deuzeman:2008sc,Hasenfratz:2013uha,Schaich:2015psa}.  Spectrum results in the weak coupling regime are consistent with  dilaton chiral perturbation theory  (dChiPT), which is designed to describe systems very close but below the conformal sill \cite{Appelquist:2017wcg,LatticeStrongDynamics:2018hun,Appelquist:2019lgk,Golterman:2020tdq,Appelquist:2020bqj}. However, dChiPT fits found that the 
extrapolation of the lattice data to the chiral limit is very long,
and  the dynamically-generated IR scale of the massless Nf=8 theory is much smaller than all the scales probed by simulations.
It is also feasible that  dChiPT  describes systems very close but above the conformal sill\cite{Appelquist:2020xua}.

 Staggered fermions are equivalent to Dirac-K\"ahler fermions and correspond to four Dirac flavors per staggered species  at the vanishing gauge coupling of the Gaussian FP \cite{Banks:1982iq,Becher:1982ud,Joos:1986cq}. The renormalized trajectory (RT) that originates from the GFP preserves this property, so $N_s$ staggered fermions describe $4N_s$ Dirac flavors around the GFP, and even at the IRFP, if the system is conformal.  Beyond the IRFP the equivalence does not have to hold. 

Most of the existing studies with 2 staggered fields found a first order bulk transition, and many identified a novel strong coupling phase
 that exhibits spontaneous breaking of the single-site shift symmetry of staggered fermions (S4 phase) \cite{Cheng:2011ic}. The S4 phase appears to be chirally symmetric and confining
 \footnote{ The spectrum in the S4 phase has been studied extensively with $N_f=12$ flavors (3 staggered fermions) in Ref. \cite{Cheng:2011ic}. In the $N_f=8$ case only the existence of the phase as evidenced by the S4 order parameter has been established. Simulations with finite fermion mass in the S4 phase are under way and publication describing the meson spectrum is in preparation.}, 
 conceivably describing  symmetric mass generation (SMG) \cite{Tong:2021phe,Butt:2021brl,Butt:2021koj}.  Two staggered fermion species in the massless limit are equivalent to four reduced staggered fields that correspond to 16 Weyl spinors. This is exactly the fermion number needed to cancel all t'Hooft anomalies, a necessary condition for the fermions to become massive while remaining chirally symmetric.  Thus the model could describe SMG if an appropriate four-fermion interaction is induced~\cite{ Butt:2021brl,Butt:2021koj}. 
 Gauge fields with momentum close to the cutoff might generate such interaction between the fermion doublers. This possibility has not been pursued since a universal continuum limit cannot be defined at first order phase transitions. The existence of a continuous phase transition would change the picture. 

The results I present here were obtained using  uniquely improved lattice actions that contain  heavy Pauli-Villars (PV) type bosonic fields \cite{Hasenfratz:2021zsl}, similar to the PV regulators in perturbation theory \cite{Pauli:1949zm}.  The mass of the PV bosons is  always kept comparable to the cutoff, thus the PV fields do not influence the IR dynamics. However, they generate a local  effective gauge action that shifts the bare gauge coupling towards weaker coupling where the UV fluctuations are suppressed. As the number of PV fields increases, or their mass decreases,  the bulk phase transition is also shifted.  Both in Ref. \cite{Hasenfratz:2021zsl} and in the present work it is observed that  the renormalized gauge coupling  increases, while the discontinuity  decreases at the phase transition as PV bosons are added. In the $N_f=8$ system the discontinuity at the phase transition  eventually disappears, suggesting a continuous  transition. This could imply that the first order transition observed without PV bosons at strong coupling is due to lattice fluctuations and opens the possibility that a non-perturbative, strongly coupled UVFP  emerges as the cutoff effects are reduced by improved  actions.

I investigate two lattice actions with different PV boson contents, in addition to the one without PV fields. I use finite size scaling to predict the critical coupling  and the critical exponent at the phase transition. In the analysis  I use the finite volume gradient flow renormalized coupling $g^2_{GF}$  \cite{Luscher:2010iy} at various flow times fixed by the lattice size $L$ as $\sqrt{8t} = c L$, where $c$ is an arbitrary parameter that I vary between 0.3 and 1.0, though the conclusion is based on the results at $c=0.45$ \cite{Fodor:2012td}.  Inspired by the arguments in Refs. \cite{Miransky:1998dh,Kaplan:2009kr,Gorbenko:2018ncu}, I   also consider a scaling form corresponding to an RG $\beta$ function that touches the axis at $g^2_\star$, i.e. $\beta(g^2) \sim (g^2-g^2_\star)^2$. This  is reminiscent to the Berezinsky, Kosterlitz, Thouless (BKT) scaling of the 2 dimensional XY model, though with exponent $\nu=1.0$, instead of $\nu=0.5$~\cite{Berezinsky:1970fr,Kosterlitz:1973xp,Kosterlitz:1974sm}.  The finite size scaling analysis for the action without PV bosons is consistent with a first order transition with critical exponent $\nu$ close to 1/4.
The  PV improved actions  show good curve collapse and consistent exponents   for  for ``walking scaling" and $\nu\approx 1.0$, though second order scaling cannot be fully excluded either. However, numerical results  of the PV improved actions are not 
consistent with a first order transition.

The rest of this paper is organized as follows. Sect. \ref{action} describes the lattice actions and the effect of the additional PV bosons. In Sect. \ref{fss-th} I outline finite size scaling with the gradient flow coupling and discuss the relevant scaling forms. Sect. \ref{transition} compares the phase structure of the PV improved actions and the original action without PV fields. Finally, I present the result of the finite size scaling analysis in Sect. \ref{fss}. Sect. \ref{summary} summarizes the  results and discusses open questions.

%%%%%%%%%%%%%%%%%%%%%%%%%%%%%%%%%%%%%%%%%%%%%%%
\section{\label{lattice} Lattice definitions} %Complete
%%%%%%%%%%%%%%%%%%%%%%%%%%%%%%%%%%%%%%%%%%%%%%%
%%%%%%%%%%%%%%%%%%%%%%%%%%%%%%%%%%%%%%%%%%%%%%%
\subsection{\label{action} The PV lattice action}%Complete
%%%%%%%%%%%%%%%%%%%%%%%%%%%%%%%%%%%%%%%%%%%%%%%

I use a gauge action that includes
fundamental and adjoint plaquette terms
with couplings $\beta\equiv\beta_F$ and $\beta_A$, 
related by $\beta_A / \beta_F = -0.25$ \cite{Cheng:2011ic}.
The gauge links in the staggered fermion operator are nHYP-smeared~\cite{Hasenfratz:2001hp, Hasenfratz:2007rf}
with smearing parameters $\alpha = (0.5, 0.5, 0.4)$.
This lattice action has been used in multiple 
8-, 4+8, and 12-flavor works~\cite{Cheng:2011ic, Hasenfratz:2013uha,Hasenfratz:2014rna,Hasenfratz:2016dou,Cheng:2013eu,Brower:2015owo,Hasenfratz:2016gut},
including the large-scale spectrum studies
of the LSD collaboration \cite{Appelquist:2016viq,Appelquist:2018yqe}. 

In the simulations I set the fermion mass to $a m_f=0$ and use symmetric $L^4$ volumes with $L/a=8$, 10, 12, 16, and 20. The fermion boundary condition is antiperiodic in all four directions to control the fermion zero modes. The gauge fields obey periodic boundary conditions. Simulations with $a m_f=0$ are well behaved in conformal systems or in the small volume deconfined regime of QCD-like systems.  The S4 phase is confining but chirally symmetric, and the pions are not particularly light \cite{Cheng:2011ic}. While numerical simulations are increasingly difficult at strong gauge coupling, the Hybrid Monte Carlo (HMC) updates are well controlled even in the chiral limit of the S4 phase.

For each staggered fermion field, I include $N_{PV}$ PV bosons. The PV bosons have the same action as the fermions but with
bosonic statistics \cite{Hasenfratz:2021zsl}. They are degenerate in mass and I choose their lattice mass $a m_{PV}$  to be much larger than any IR scale of the system. When  integrated out, the PV fields  generate  an effective  gauge action. This is not an ultra-local action, but still local,  with localization governed by the mass of the PV ``pions". With heavy PV mass the bound state meson masses are approximately $2 a m_{PV}$,  i.e. the range of the effective gauge action is a few lattice spacings if $2 a m_{PV} \gtrsim 1.0$. The leading term of the induced gauge action is a plaquette built from smeared links with coefficient
\be
\label{bind}
\beta^{(p)}_\text{ind} = - \frac{N_s N_{PV}}{(2 a m_{PV})^4} \ ,
\ee
where $N_s=2$ is the number of staggered fermion fields. Since $\beta^{(p)}_\text{ind}$ is negative, the bare gauge coupling $\beta$ has to increase to keep  the effective gauge coupling $\beta_\text{eff} = \beta + \beta_\text{ind}$ and  the lattice spacing constant.  In general, more PV bosons and/or smaller $a m_{PV}$ values generate a larger induced gauge action. The constant physics regime is pushed to weaker bare gauge coupling, which reduces the ultraviolet fluctuations as  evidenced by larger plaquette expectation values. The effect of the PV bosons was investigated in detail in Ref. \cite{Hasenfratz:2021zsl} for the 12-flavor SU(3) system.  I want to  emphasize that the effect of the PV bosons in the simulations is the generation of a local effective gauge action. The corresponding system has the same infrared properties at the Gaussian, as well as other possible fixed points as the original action, as long as $am_{PV} \gg 1/L$ and $am_{PV}\gg am_f$.

%%%%%%%%%%%%%%%%%%%%%%%%%%%%%%%%%%%%%%%%%%%%%%%
\subsection{\label{fss-th} Finite size scaling with gradient flow}%Complete
%%%%%%%%%%%%%%%%%%%%%%%%%%%%%%%%%%%%%%%%%%%%%%%

The nonperturbative UVFP in the 8-flavor SU(3) gauge system, if it exists, has at least two relevant parameters, the fermion mass and the new operator emerging at the UVFP. The scaling forms I consider below assumes the simplest scenario, i.e. that there are exactly two relevant parameters. If the fermion mass is set to zero, the RG flow is restricted to the critical surface of the massless theory. In that case, the UVFP has only one relevant operator, and finite size scaling methods can be used to study its properties. This is the scenario depicted in the sketches of Fig. \ref{fig:phasediag}.

Finite size scaling studies usually consider susceptibilities or bound state  masses obtained from 2-point functions. Here I  use the  gradient flow coupling $g^2_{GF}$. 
The gradient flow (GF) \cite{Narayanan:2006rf,Luscher:2009eq}.  is a smoothing transformation that has been used extensively in lattice studies  GF is not an RG transformation as it lacks the essential coarse-graining step. However,  coarse-graining can be implemented at the level of expectation values.  If one relates the lattice GF time $ t/a^2$ to the RG scale change as $b \propto \sqrt{t/a^2}$,  at large $t/a^2 \gg 1$ the GF can be interpreted as a well-defined real-space RG transformation \cite{Carosso:2018bmz}. Observables at finite flow time correspond to RG flowed observables at energy scale $\mu \propto 1/\sqrt{8 t}$, and exhibit hyperscaling in the vicinity of an RG fixed point.

 The GF coupling was proposed in Ref. \cite{Luscher:2010iy} for scale setting and in Ref. \cite{Fodor:2012td} to calculate the finite volume step scaling function in QCD-like or conformal system. It is defined as
\be
\label{GF}
\gGF(\beta,L; t) = \mathcal{N} t^2 \svev{ E(t)}_{\beta,L/a},
\ee
where $E(t)$ is the energy density at bare gauge coupling $\beta$ and flow time $t$ on lattice volume $(L/a)^4$ \footnote{If $\mathcal{N} =128\pi^2/(3(N_c^2-1))$, where $N_c$ refers to color, $\gGF$ matches the ${\overline {MS}}$ coupling at tree level perturbation theory \cite{Luscher:2010iy}. In the non-perturbative phase this normalization is arbitrary, but I keep it for consistency.}.  In lattice calculations various local operators like the plaquette or the clover operator can be used to approximate $E$. A  frequent choice to estimate $E$ is
\bea
\svev{ E}_{\beta,L/a} = 12\left(3 -  \text{Re\,tr}\langle U_\Box \rangle_{\beta,L/a}\right) \,,\\
\svev{ E(t)}_{\beta,L/a} = 12\left(3 -   \text{Re\,tr}\langle V_\Box(t) \rangle_{\beta,L/a}\right) \, ,
\eea
where $ \text{Re\,tr}\langle U_\Box \rangle$ and $ \text{Re\,tr}\langle V_\Box(t) \rangle$  denote the expectation value of the plaquette before GF flow  and  at flow time $t$. The GF defines a non-linear RG transformation, therefore $\svev{ E(t)}_{\beta,L/a}$ does not have an anomalous dimension, i.e. the $\eta$ exponent of the RG does not enter. 
Thus, the scaling dimension of the gradient flow coupling  vanishes, its RG evolution depends only on the RG flow in the action parameter space.\footnote{The use of $ t^2(3-\langle  \text{Re\,tr}\langle V_\Box(t) \rangle)$ to approximate $g^2_{GF} $ is reminiscent of the operators used in Monte Carlo Renormalization Group (MCRG) studies \cite{Swendsen:1979gn,Pawley:1984et,Bowler:1984hv,Hasenfratz:2011xn}. The main difference is that GF is continuous, so one is not restricted to scale change of 2. Also, the lattice volume does not decrease during GF, so larger flow time/ RG scale change is possible. }. This property greatly simplifies the RG scaling relations, as only the scaling exponent related to the correlation length enters.

One can think of $g^2_{GF}$ as a scalar quantity that tracks the renormalized trajectory \cite{Hasenfratz:2019hpg,Hasenfratz:2019puu}. It has several advantages, most important is that it is a simple expectation value  that is easy to measure with high precision. 
Finite size scaling relations for $g^2_{GF}$ are derived in the usual way. 
 At a second order phase transition  one expects the correlation length to scale as
\be
 \xi \propto |\beta /\beta_\star -1|^{-\nu},
\label{eq:xi_2nd}
\ee
where  $\beta_\star$ denotes the critical coupling. The combination
\be
\label{eq:scalingx}
( \beta/\beta_\star - 1 ) \, L^{1/\nu}  \propto  (\xi/L)^{-1/\nu}
\ee
is a dimensionless scaling variable.
Renormalization group scaling in finite volume therefore implies that  $\gGF(\beta,L;c)$, $c=\sqrt{8t}/L$, depends on  $( \beta /\beta_\star - 1) \, L^{1/\nu} $, i.e. follows a unique but a priori unknown scaling function 
\be
\label{eq:scaling1}
\gGF(\beta,L; c) = f_{2nd}^{(c)}(( \beta/\beta_\star -1 )  \, L^{1/\nu} ).
\ee
First order phase transitions are expected to show similar scaling behavior but with discontinuity exponent $\nu=1/d=0.25$  in 4 dimensions.
 
 If  the RG $\beta$-function behaves as
 \be
 \label{eq:beta_fn}
 \beta_{RG} \propto |\beta / \beta_\star - 1 |^{(\nu+1)} \, , \quad \nu>0 
 \ee
 in the vicinity of the critical point, the phase transition is continuous but ``walking" or BKT-type. The correlation length scales as
\be
 \xi \propto e^{\zeta |\beta /\beta_\star - 1|^{-\nu}}, \quad \quad \beta<\beta_\star
\label{eq:xi_BKT}
\ee
and the corresponding scaling  function depends on $L \, \text{exp}(-\zeta |\beta/\beta_\star-1|^{-\nu} )$, 
\be
\label{eq:scaling2}
\gGF(\beta,L; c) = f_{BKT}^{(c)} \left( L \, \text{exp}(-\zeta |\beta/\beta_\star - 1|^{-\nu}) \right).
\ee
The 2-dimensional XY model has $\nu=1/2$, while the quadratic ``walking"  $\beta$ function in Eq. \ref{eq:beta_fn} corresponds to $\nu=1$.   Both Eqs. \ref{eq:scaling1} and  \ref{eq:scaling2} are leading order relations valid only in the vicinity of the critical point, $|\beta-\beta_\star|\ll 1$.  One also has to assure that  the flow does not drive the  RG blocked correlation length below the lattice spacing,  $\sqrt{t/a^2} \lesssim \xi$. At the same time $\sqrt{t/a^2}$ has to be large enough for the RG flow to reach the renormalized trajectory. This latter condition could limit the use of the smallest volumes, since $t=(c L)^2/8$ and  $c\le 0.5$ is commonly preferred\footnote{At larger $c$ values the GF ``wraps around" the finite volume. That does not invalidate the use of $g^2_{GF}$, though the statistical errors increase with increasing $c$. A full investigation on the dependence on $c$, similar to Ref.\cite{Fritzsch:2013je}, is a worthwhile project for the future.} . The RT is approached  according to the largest non-leading exponent,  but corrections to scaling could be important. This, however, is beyond what my present data set can describe.

The scaling functions $f_{2nd}^{(c)}$  and $f_{BKT}^{(c)}$ depend on the parameter $c$, but  should not depend on $\nu$,  $\zeta$, and $\beta_\star$. In addition, different lattice actions should show scaling with the same  $\nu$ exponent, though $\beta_\star$ and  $\zeta$ are dependent on the action. Thus testing different $c$ values and different actions provides a consistency check  and verifies the scaling form.

%%%%%%%%%%%%%%%%%%%%%%%%%%%%%%%%%%%%%%%%%%%%%%%
\section{\label{results} Numerical results}
%%%%%%%%%%%%%%%%%%%%%%%%%%%%%%%%%%%%%%%%%%%%%%%

%%%%%%%%%%%%%%%%%%%%%%%%%%%%%%%%%%%%%%%%%%%%%%%
\subsection{\label{transition} The phase structure}
%%%%%%%%%%%%%%%%%%%%%%%%%%%%%%%%%%%%%%%%%%%%%%%

%%%%%%%%%%%%%%%%%%%%%%%%%%%%%%%%%%%%%%%%%%%%%%%%%%%%%%%%%%%%%%%%%%%%%%%%%
\begin{figure}[t]
\vspace*{-2ex}
%\hspace{-15mm}
\includegraphics[width=0.95\columnwidth]{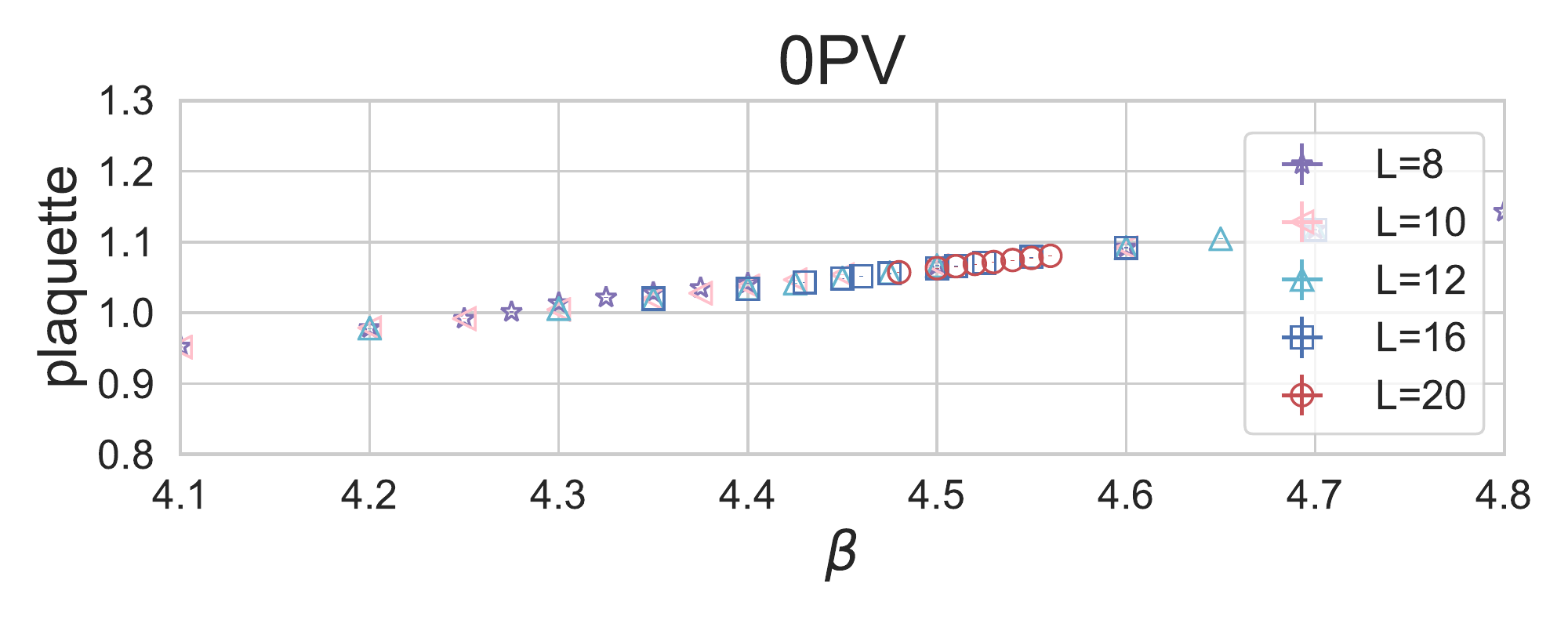} \\
\includegraphics[width=0.95\columnwidth]{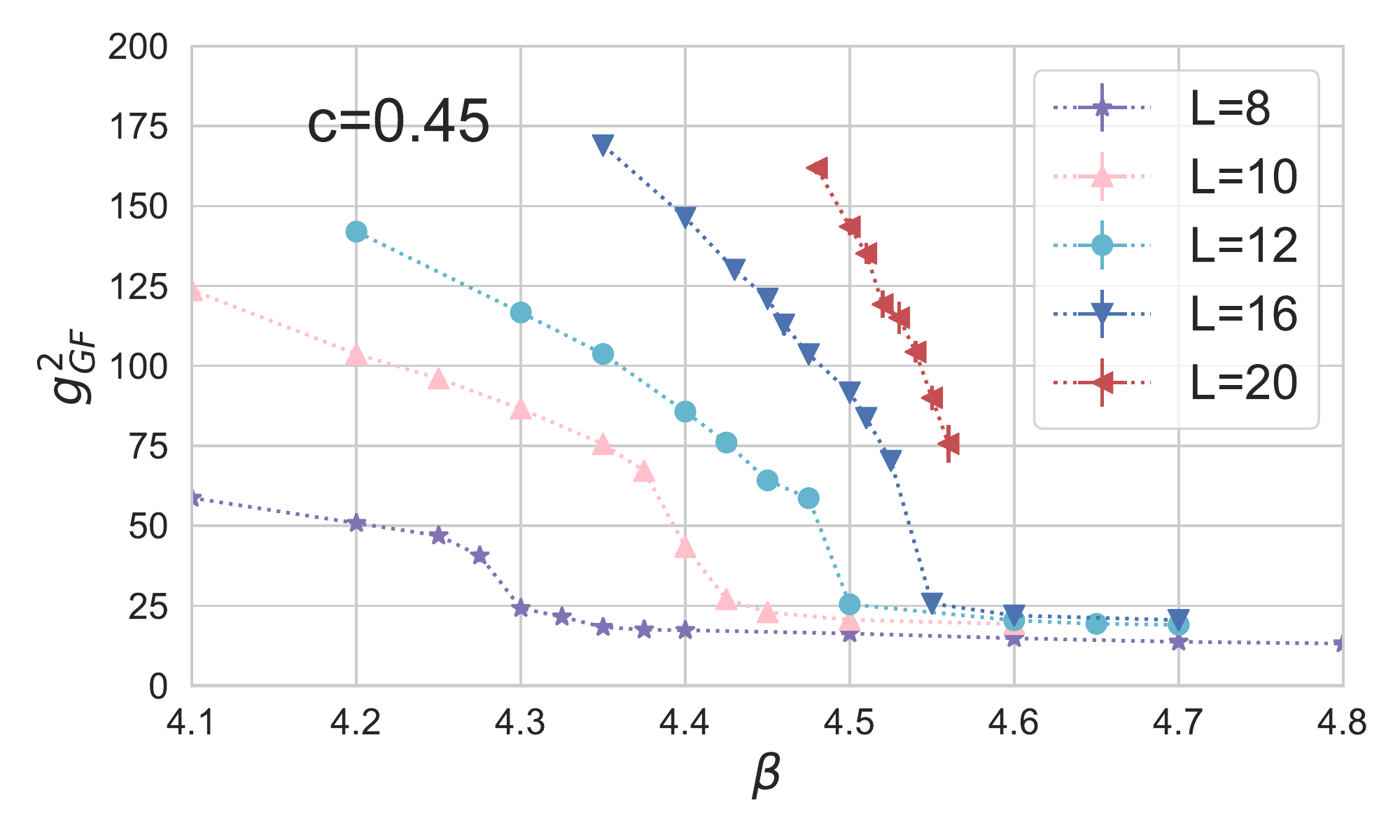} \\
\includegraphics[width=0.95\columnwidth]{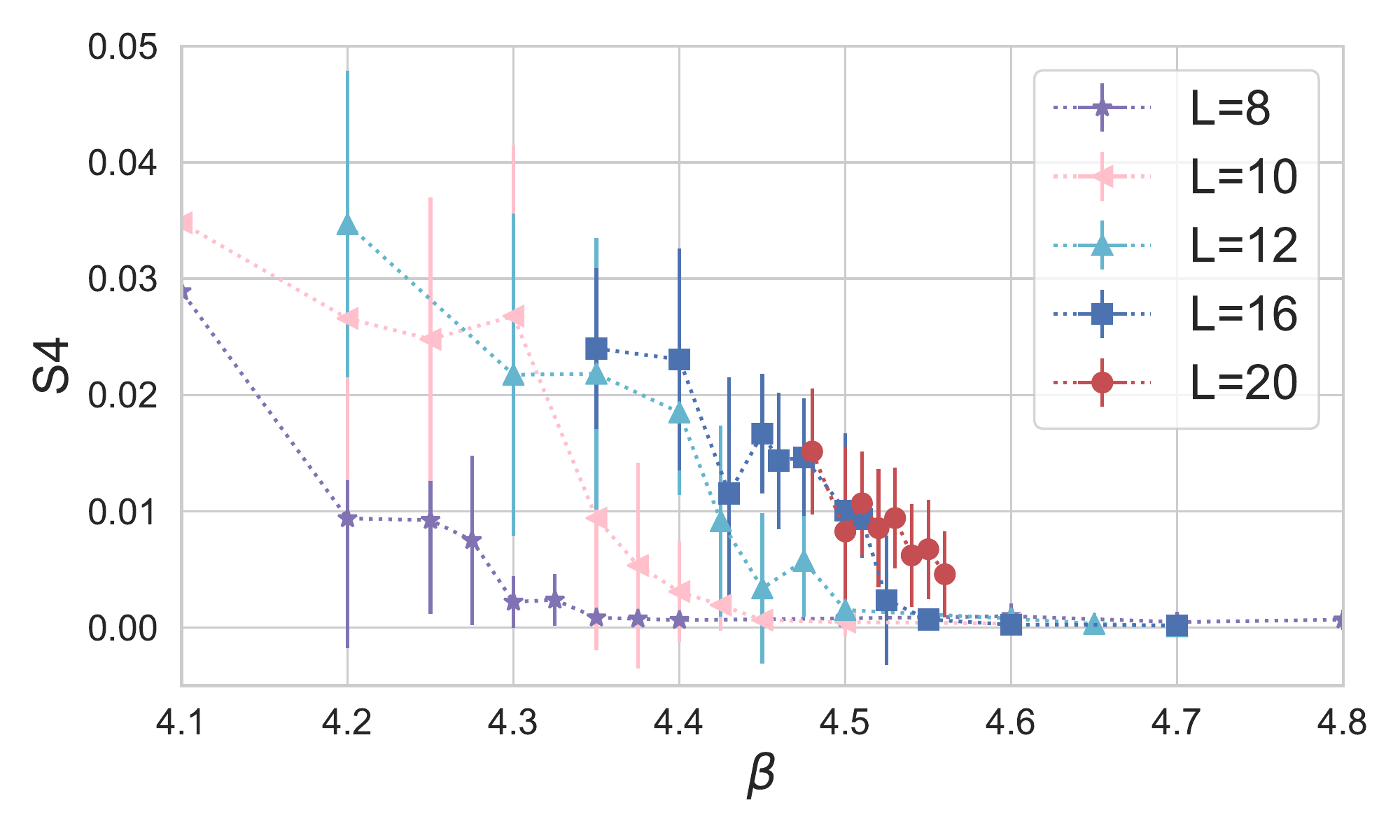} \\
\includegraphics[width=0.95\columnwidth]{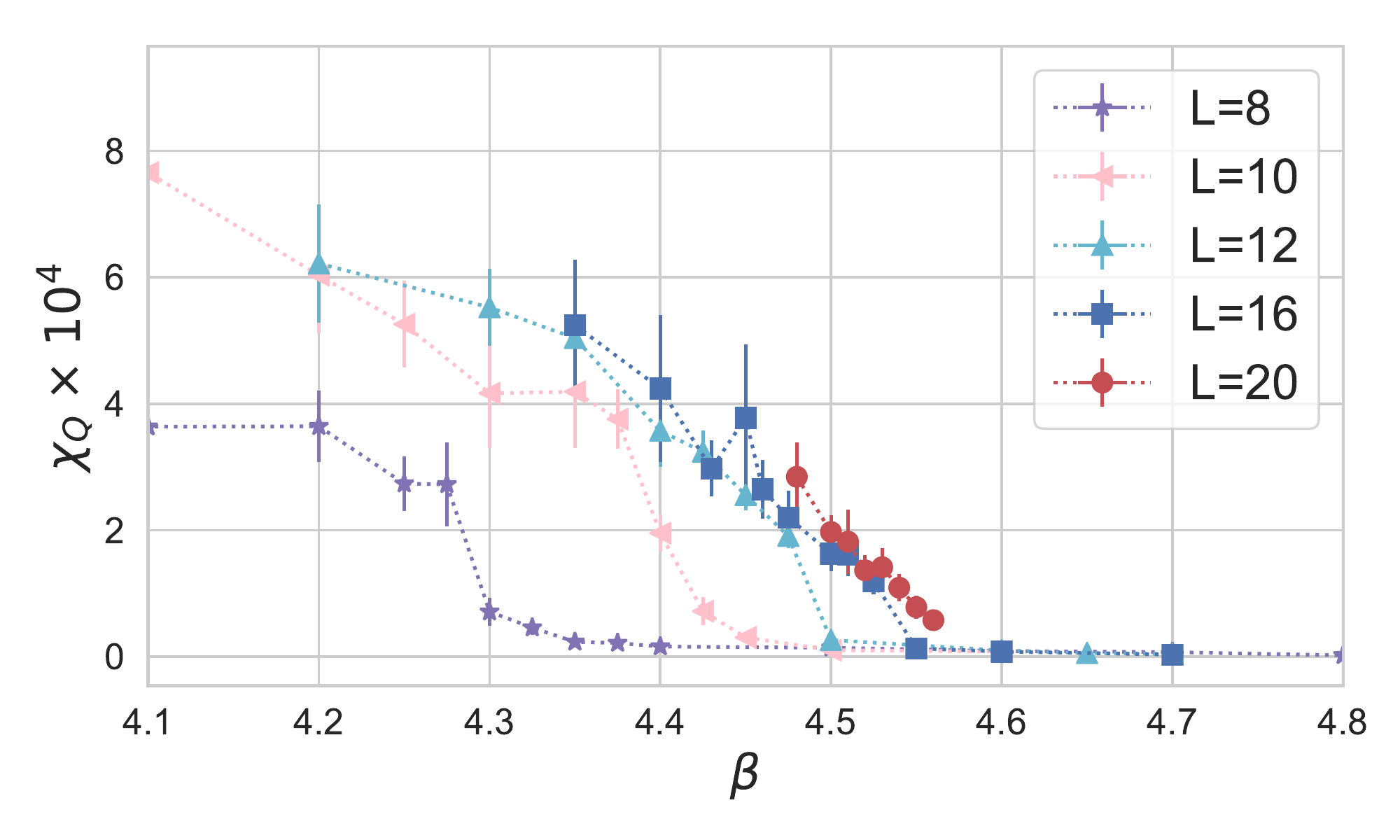} 
\vspace{-2ex}
\floatcaption{fig:0PV}%
{
The plaquette, the gradient flow coupling $\gGF$ at $c=0.45$, the S4 order parameter, and the topological susceptibility as the function of the bare gauge coupling $\beta$ for the  0PV system. All simulations were done in the chiral limit on  $L^4$ volumes. The dashed lines are guides to the eye.
}
 \end{figure}
%%%%%%%%%%%%%%%%%%%%%%%%%%%%%%%%%%%%%%%%%%%%%%%%%%%%%%%%%%%%%%%%%%%%%%%%%

%%%%%%%%%%%%%%%%%%%%%%%%%%%%%%%%%%%%%%%%%%%%%%%%%%%%%%%%%%%%%%%%%%%%%%%%%
\begin{figure}[t]
\vspace*{-2ex}
%\hspace{-15mm}
\includegraphics[width=0.95\columnwidth]{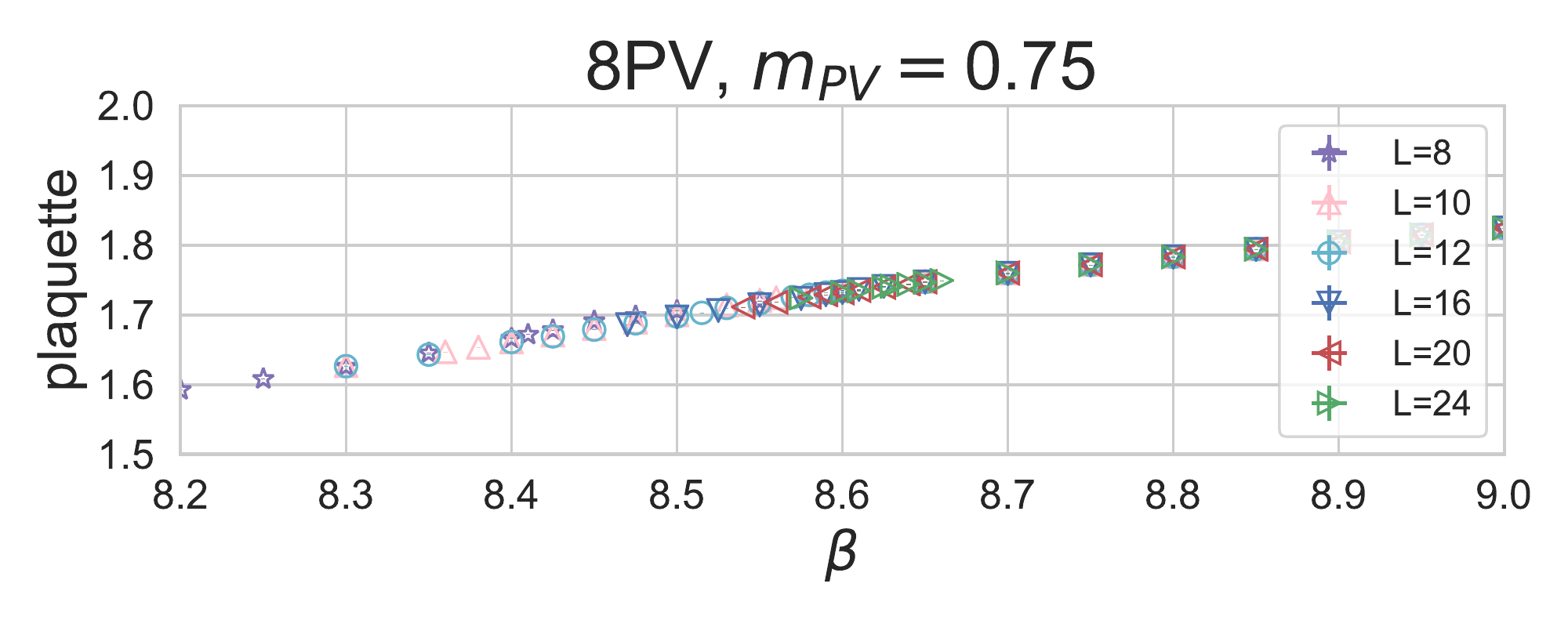} \\
\includegraphics[width=0.95\columnwidth]{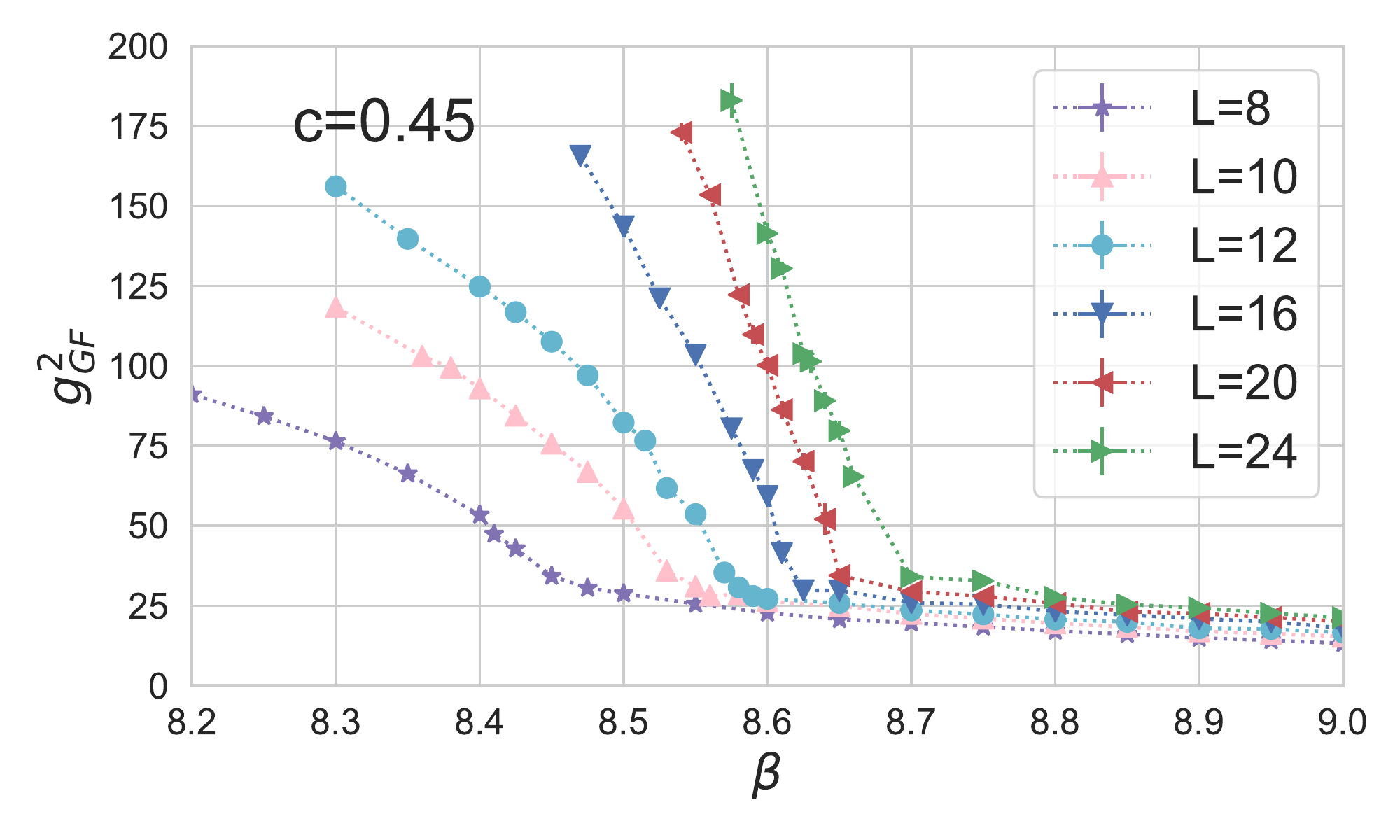} \\
\includegraphics[width=0.95\columnwidth]{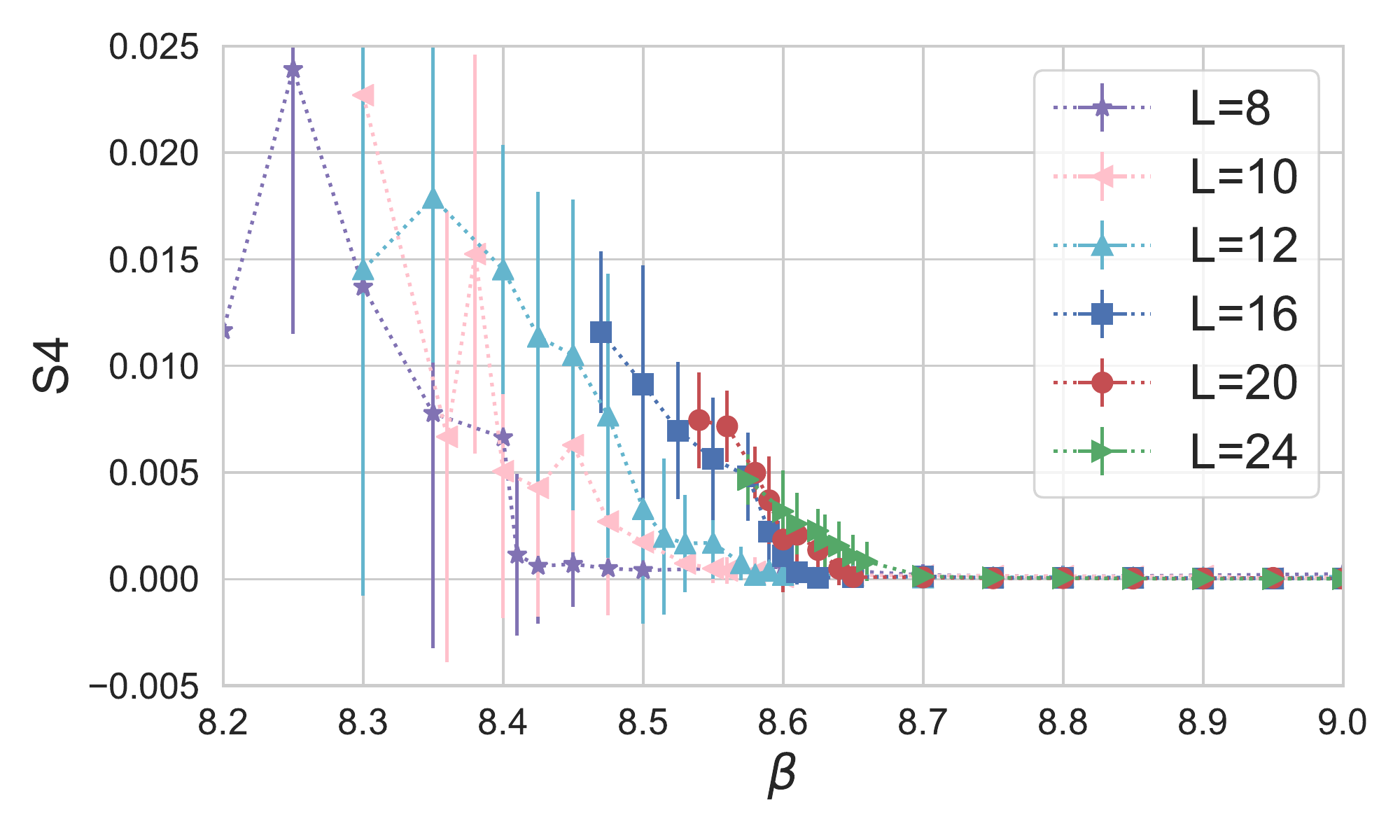} \\
\includegraphics[width=0.95\columnwidth]{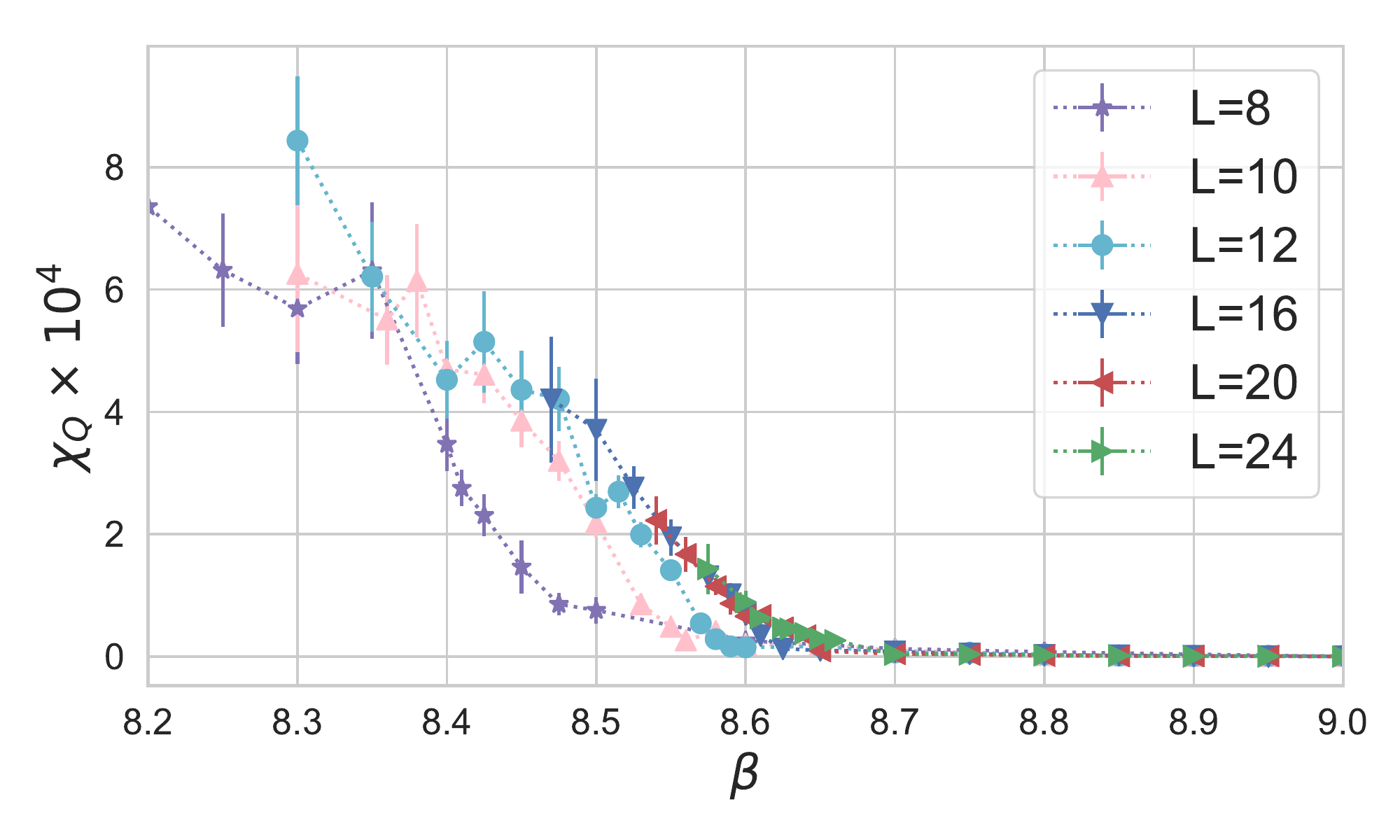} 
\vspace{-2ex}
\floatcaption{fig:8PV-m0.75}
{
Same as Fig. \ref{fig:0PV} but for the 8PV-m0.75 action.
}
 \end{figure}
 %%%%%%%%%%%%%%%%%%%%%%%%%%%%%%%%%%%%%%%%%%%%%%%%%%%%%%%%%%%%%%%%%%%%%%%%%
%%%%%%%%%%%%%%%%%%%%%%%%%%%%%%%%%%%%%%%%%%%%%%%%%%%%%%%%%%%%%%%%%%%%%%%%%
\begin{figure}[t]
\vspace*{-2ex}
%\hspace{-15mm}
\includegraphics[width=0.95\columnwidth]{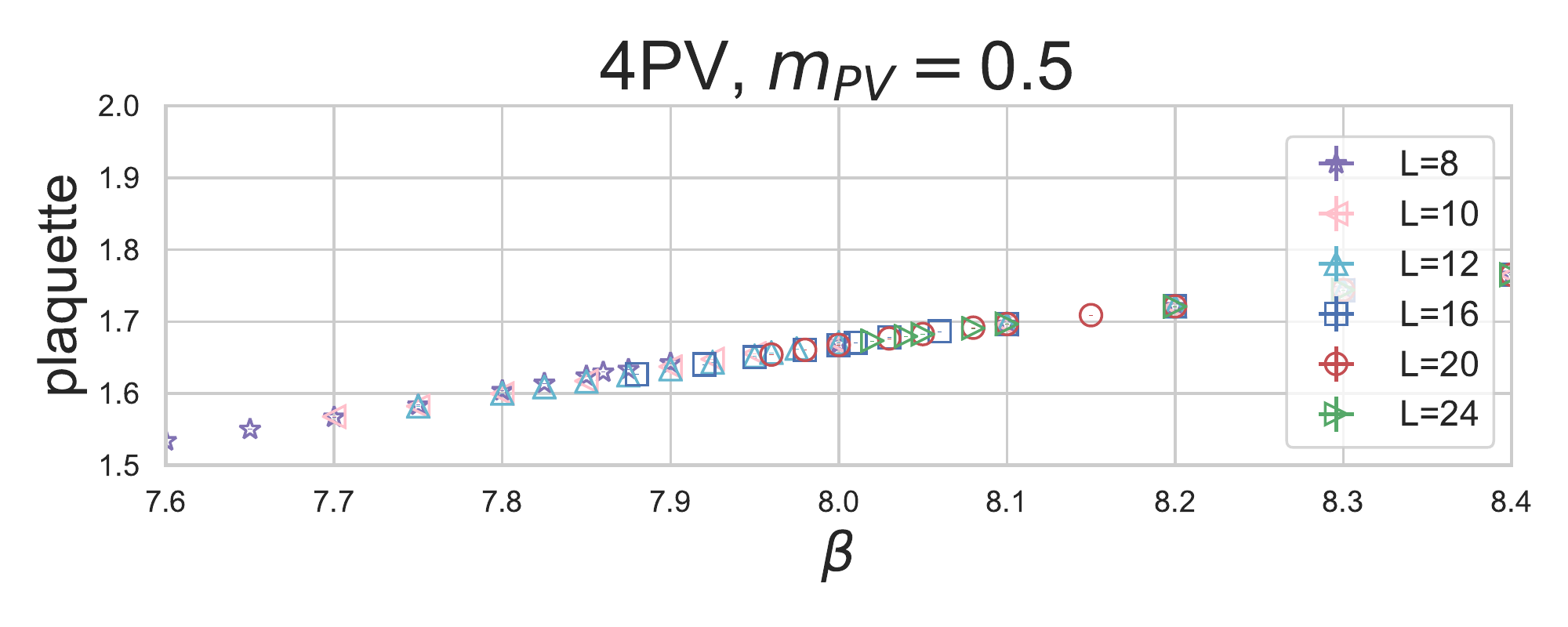} \\
\includegraphics[width=0.95\columnwidth]{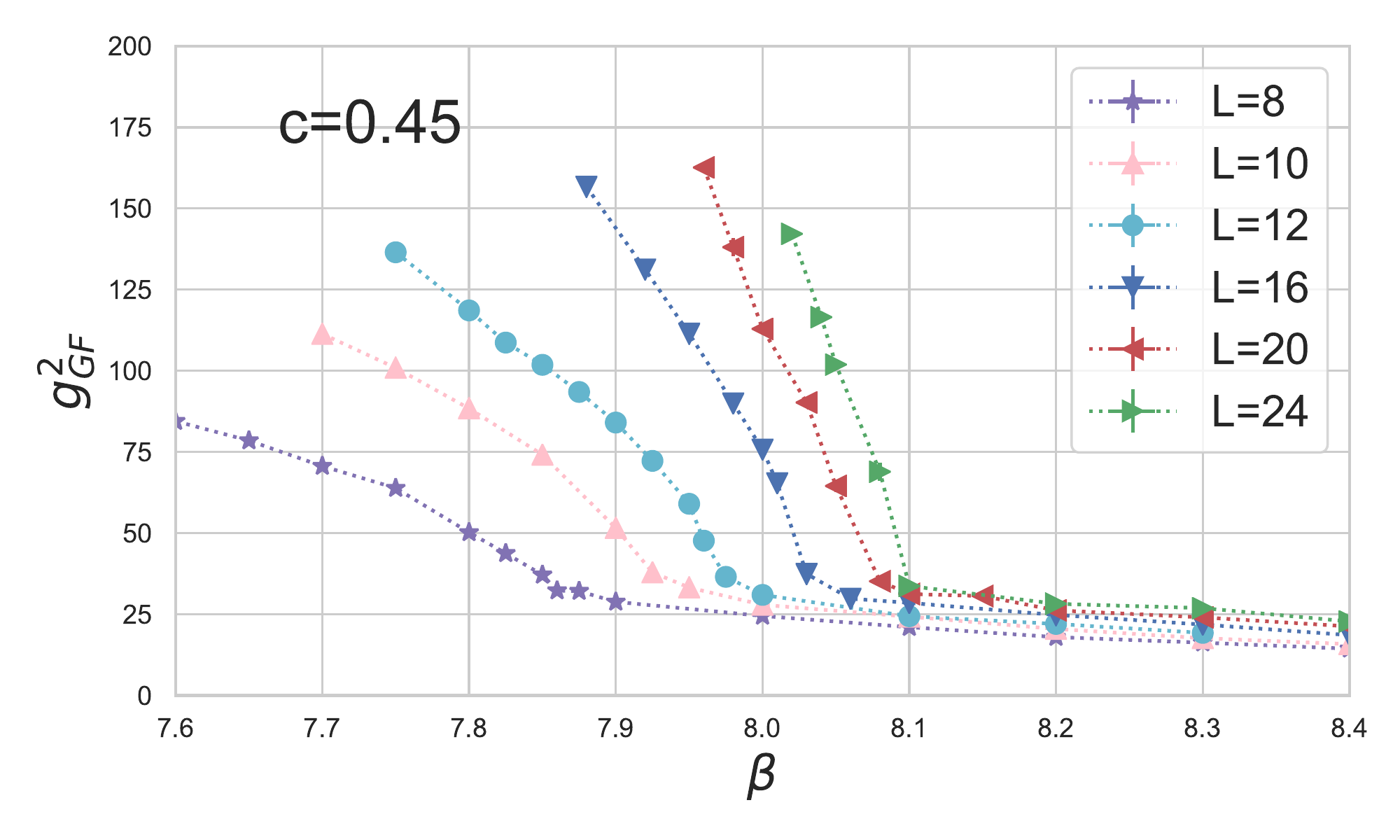} \\
\includegraphics[width=0.95\columnwidth]{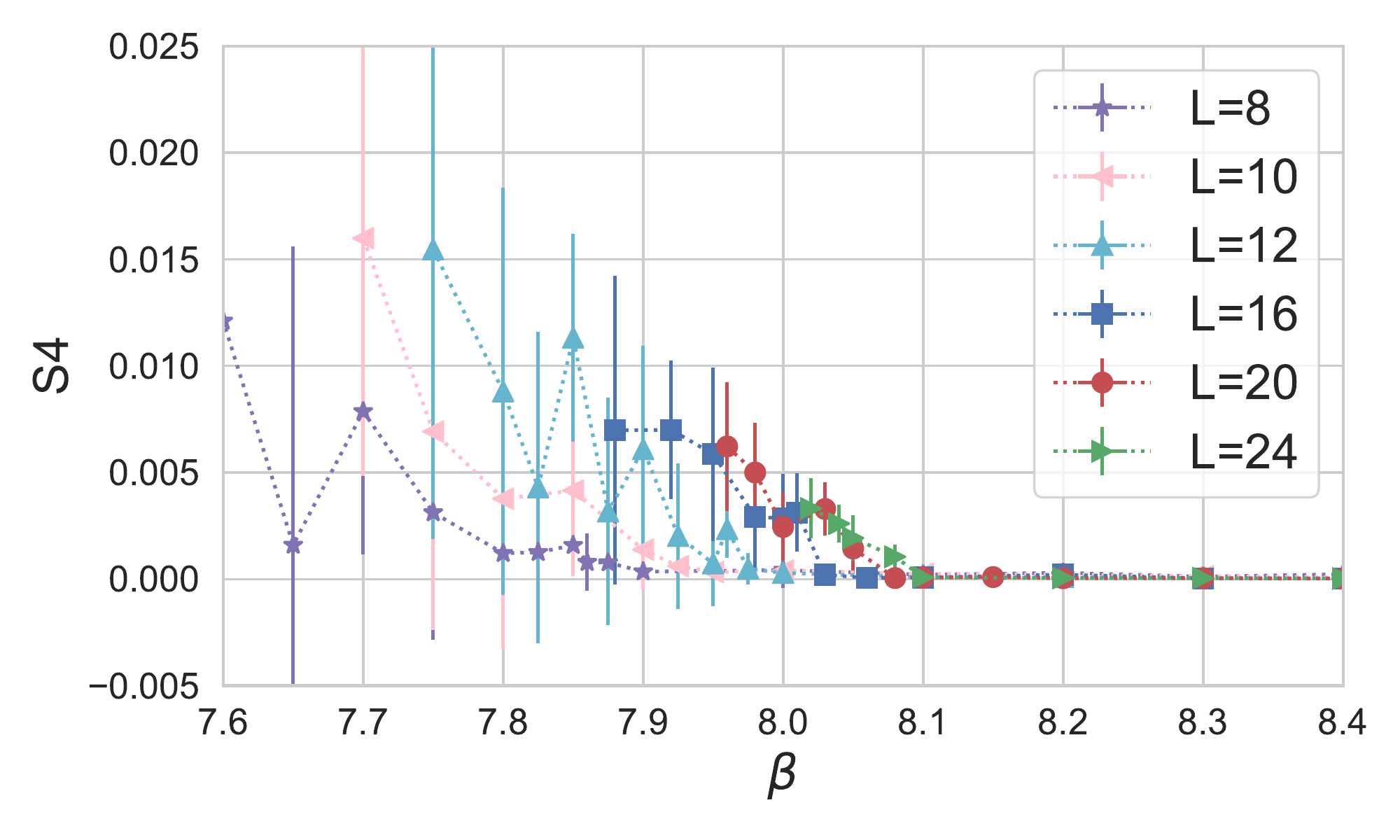}\\
\includegraphics[width=0.95\columnwidth]{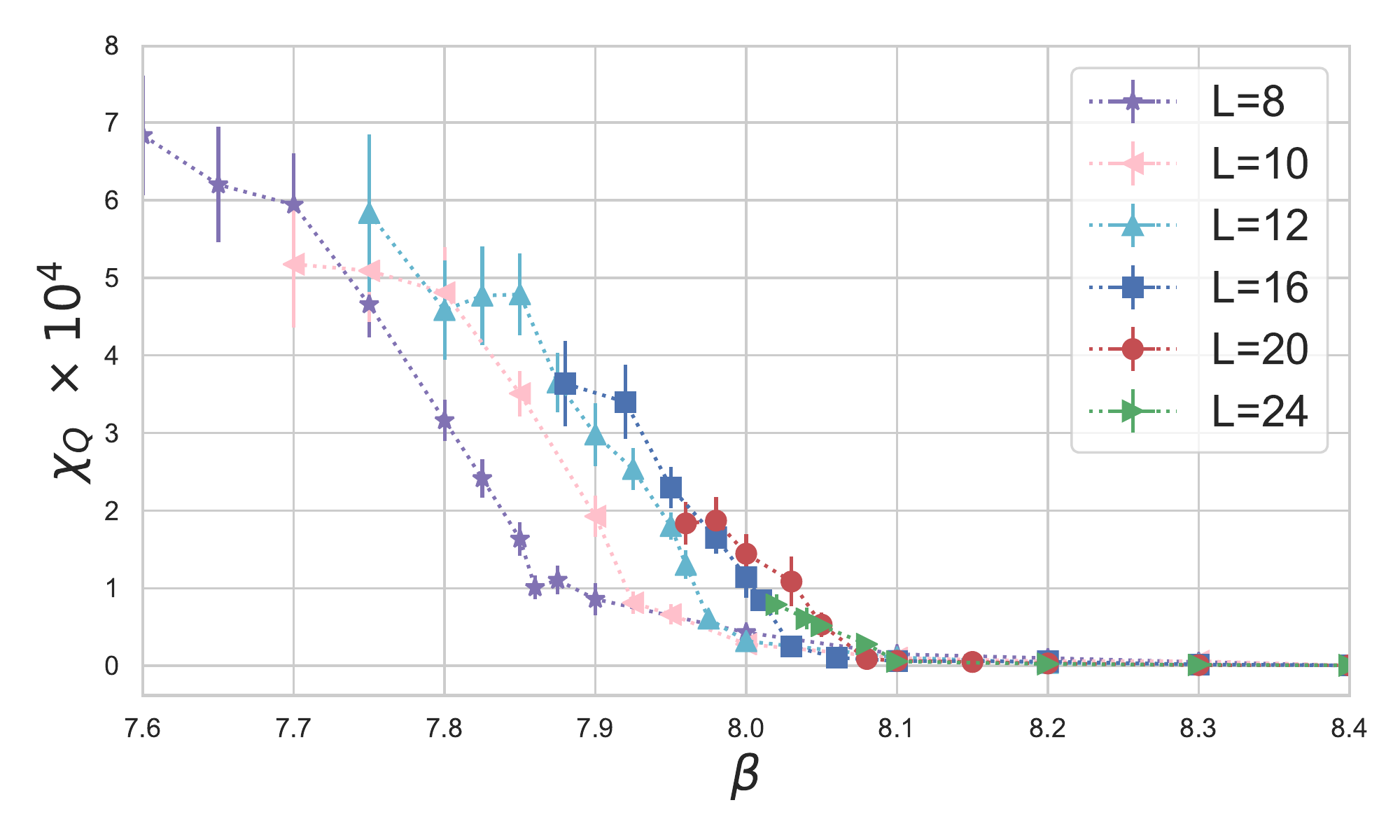} 
\vspace{-2ex}
\floatcaption{fig:4PV-m0.5}
{
Same as Fig. \ref{fig:8PV-m0.75} but for the 4PV-m0.5 action.
}
 \end{figure}
%%%%%%%%%%%%%%%%%%%%%%%%%%%%%%%%%%%%%%%%%%%%%%%%%%%%%%%%%%%%%%%%%%%%%%%%%

I investigated several different PV boson actions, two of them in detail. The first one contains 8 PV fields for each staggered flavor with mass $a m_{PV}=0.75$ (8PV-m0.75 action), and  the second one contains 4 PV fields  per staggered flavor  with mass $a m_{PV}=0.5$ (4PV-m0.5 action). 
I have also looked at the action without PV bosons (0PV action).
The 0PV action was studied previously in Ref. \cite{Hasenfratz:2013uha} at small but finite mass, and in the chiral limit but at weaker couplings in Ref. \cite{Hasenfratz:2014rna}.  No previous work considered  the 8-flavor systems at the phase transition directly in the chiral limit. In contrast, all simulations I present here are performed with $am_f=0$.

I start by comparing the plaquette expectation value, the finite volume GF coupling, the S4 order parameter for the plaquette, and the topological susceptibility of the three actions.
The top panel of Fig. \ref{fig:0PV} shows the plaquette as the function of the bare gauge coupling $\beta$ on volumes $L/a=8$, 10, 12, 16, and 20  for the 0PV action. The discontinuity at the phase transition is too small to be resolved on the plot.
 Nevertheless, I have observed tunneling  and 2-state signals in several ensembles, consistent with a first order phase transition.
 The  GF coupling $\gGF$ at $c\equiv\sqrt{8t}/L=0.45$ is shown on the second panel. It exhibits a rapid change around $\beta \approx 4.3 - 4.55 $ on these volumes, consistent with a phase transition  around $\beta_{1st}\approx 4.6$, the value predicted with small but finite fermion masses in Ref.\cite{Hasenfratz:2013uha}. 
 The strong coupling phase exhibits spontaneous breaking of the staggered single-site shift symmetry (S4 phase). The S4 order parameter  that measures the staggered nature of the plaquette (Eq. 3. in Ref. \cite{Cheng:2011ic}) is shown on the third panel. 
 It is  fairly noisy but  useful to verify that the new phase shows the S4 symmetry breaking pattern. 
 The S4 phase is confining but chirally symmetric, and the existence of a local  order parameter ensures that it is separated from both  the conformal and  the chirally broken, confining phases by a phase transition \cite{Cheng:2011ic} 
 The bottom panel of  Fig. \ref{fig:0PV} shows the topological susceptibility $\chi_Q =\langle Q^2\rangle/L^4$. In the chiral $a m_f=0$ limit of conformal and QCD-like systems one expects  $\chi_Q=0$, because the fermion determinant has a zero mode on isolated instantons. Staggered fermion lattice artifacts lift $ \chi_Q$ slightly at finite lattice spacing \cite{Follana:2004sz}. However, in the S4 phase, the topological charge is large and no longer suppressed.
 This is  further evidence of the very different nature of this new phase. 
 
 The largest GF coupling in the weak coupling phase of the 0PV model is  $\gGF\lesssim 26$ with $L=16$ and $c=0.45$.
 The plaquette is fairly small at the phase transition, $ \text{Re\,tr}\langle U_\Box \rangle  \approx 1.0$ (normalized to 3.0), suggesting large UV fluctuations.  This system has another first order phase transition at an even stronger coupling that separates the S4 phase from a QCD-like chirally broken, confining regime \cite{Hasenfratz:2013uha}. I am not concerned with that transition in this work.

The  panels of Figs. \ref{fig:8PV-m0.75} and \ref{fig:4PV-m0.5} show the same quantities as Fig. \ref{fig:0PV}  but for the   8PV-m0.75 and 4PV-m0.5  actions and volumes $L\le 24$~\footnote{Simulations with the PV actions are computationally faster, similar to the observation  presented in Table I of Ref.~\cite{Hasenfratz:2021zsl} for $N_f=12$, thus larger volumes are feasible.}. The GF coupling now appears continuous. None of my numerical simulations showed  2-state signals or phase tunneling either. 
The S4 order parameter becomes non-zero  at the same place where
$\gGF$ shows a qualitative change of $\beta$ dependence,  around $\gGF \gtrsim 35$, $c=0.45$. This is also where the S4 order parameter and topological susceptibility show a sudden increase. The plaquette is significantly larger at the phase transition, $ \text{Re\,tr}\langle U_\Box \rangle \approx 1.75$ and $ \text{Re\,tr}\langle U_\Box \rangle \approx 1.7$, respectively. The bare gauge coupling of the phase transition has also shifted to larger bare coupling, from $\beta_\star \approx 4.6$ of the 0PV action to $\beta_\star\approx 8.7$ and $\beta_\star\approx 8.1$. This is the consequence of the induced effective action.

The strongest GF coupling with the 0PV action  is $\gGF\approx 26$ on the volumes considered, compared to $\gGF \approx 35$ with both PV actions. The 0PV action has large vacuum fluctuations. It appears   that those trigger a first order phase transition before the GF coupling is strong enough to undergo the  continuous transition seen with the 4PV-m0.5 and 8PV-m0.75 actions. This suggests that the first order phase transition  could be only a lattice artifact. The picture is similar to the frequently observed first order bulk transition with Wilson fermions that are not considered physical.

%%%%%%%%%%%%%%%%%%%%%%%%%%%%%%%%%%%%%%%%%%%%%%%%%%%%%%%%%%%%%%%%%%%%%%%%%
\begin{figure*}[t]
%\vspace{-1ex}
%\hspace{-15mm}
\includegraphics[width=0.99\columnwidth]{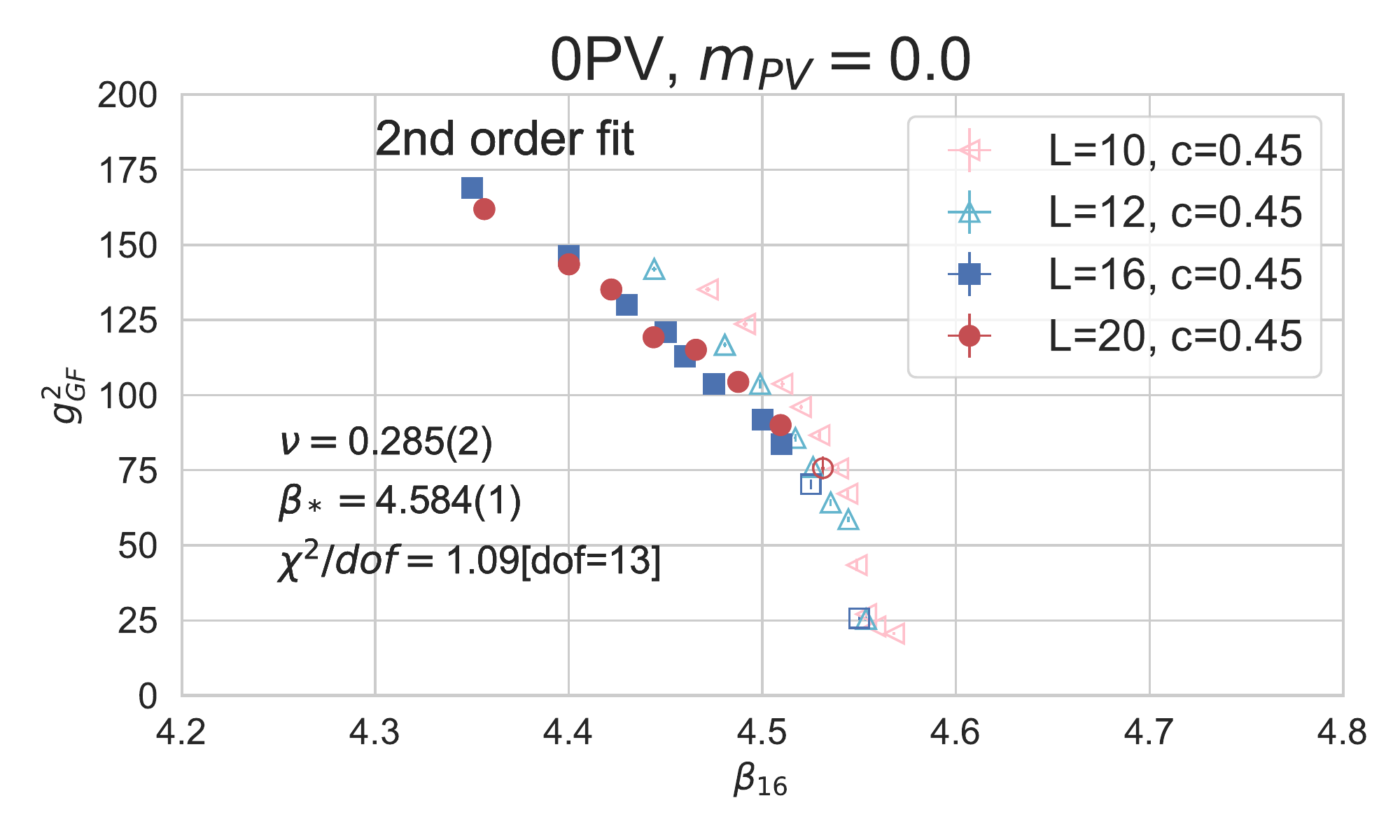} 
\includegraphics[width=0.99\columnwidth]{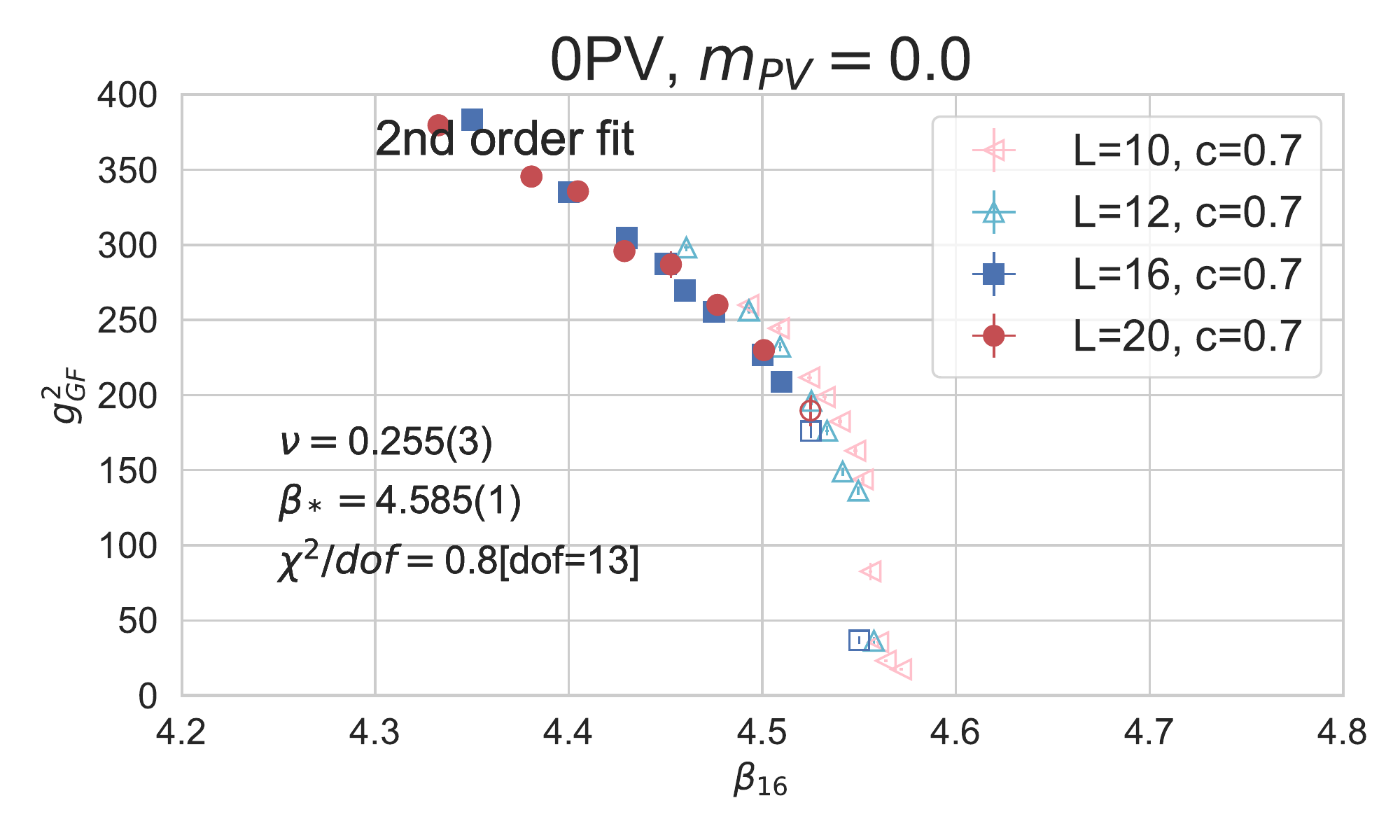} \\
\includegraphics[width=0.99\columnwidth]{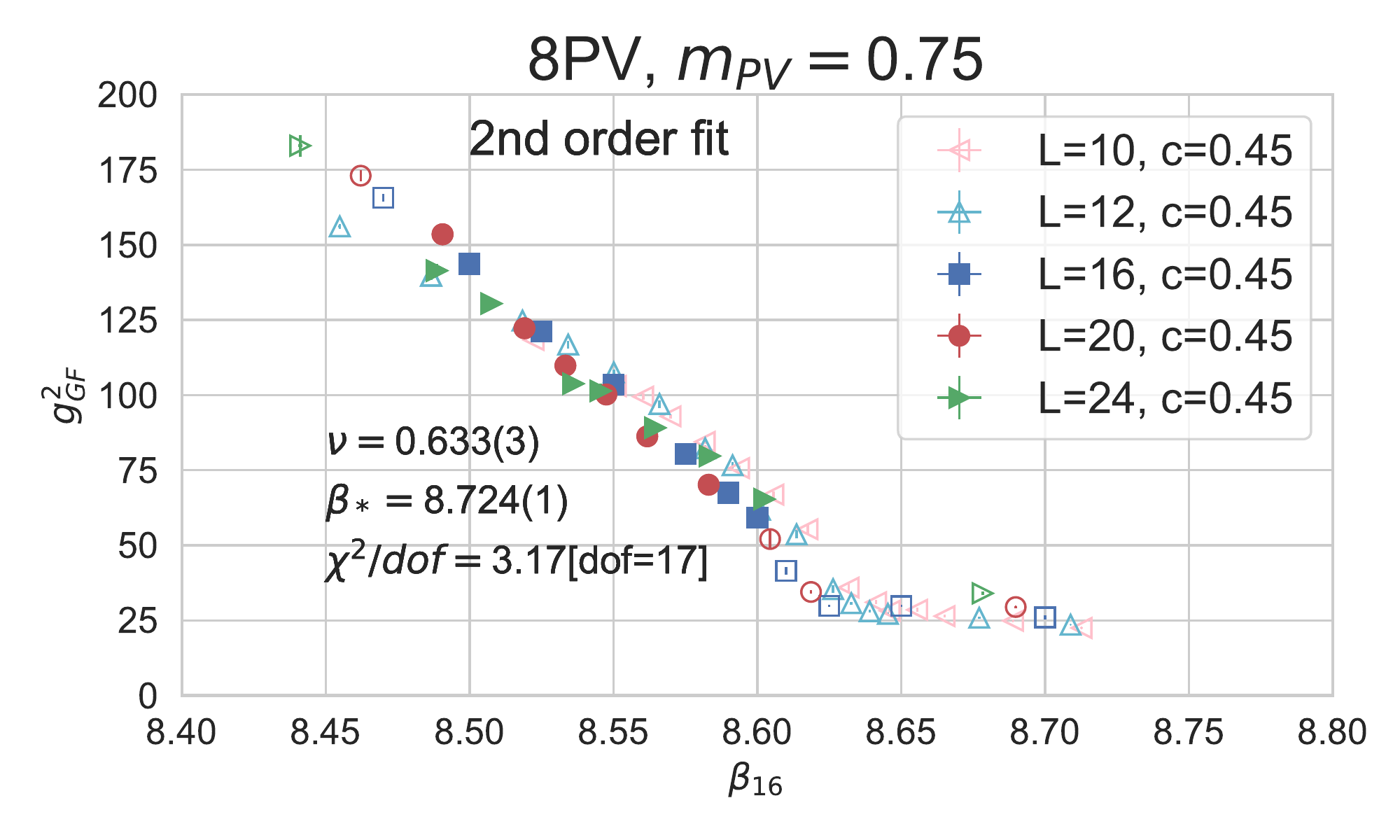} 
\includegraphics[width=0.99\columnwidth]{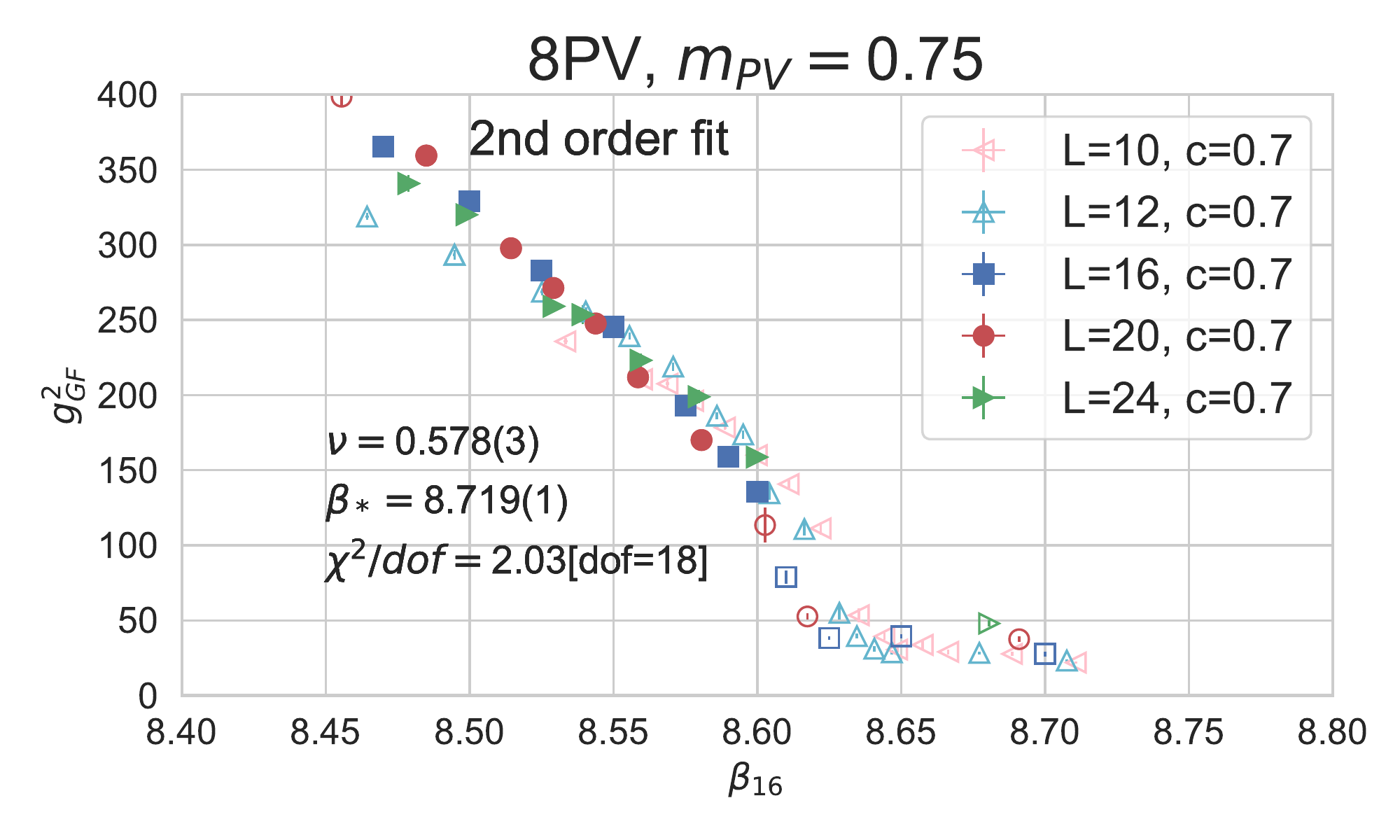} \\
\includegraphics[width=0.99\columnwidth]{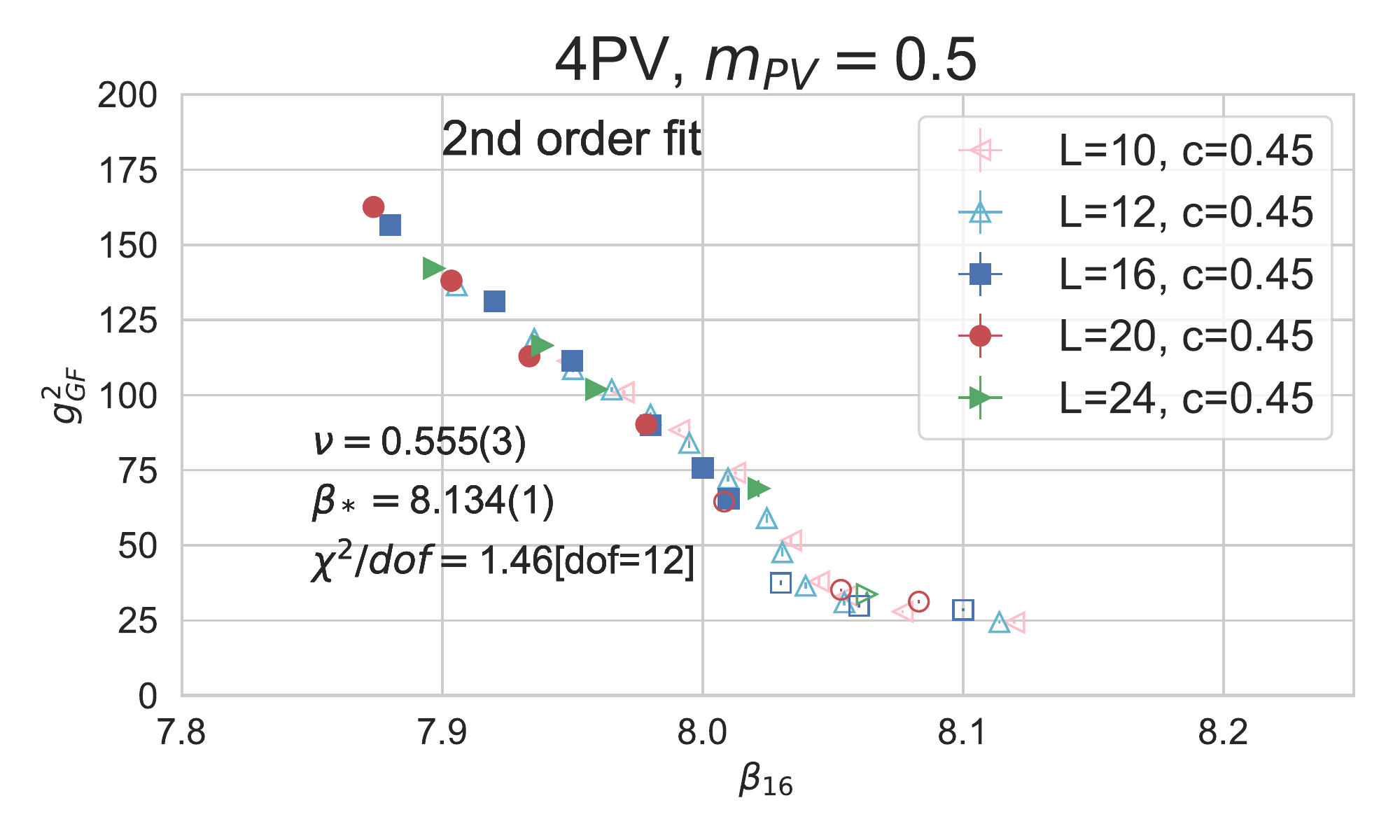} 
\includegraphics[width=0.99\columnwidth]{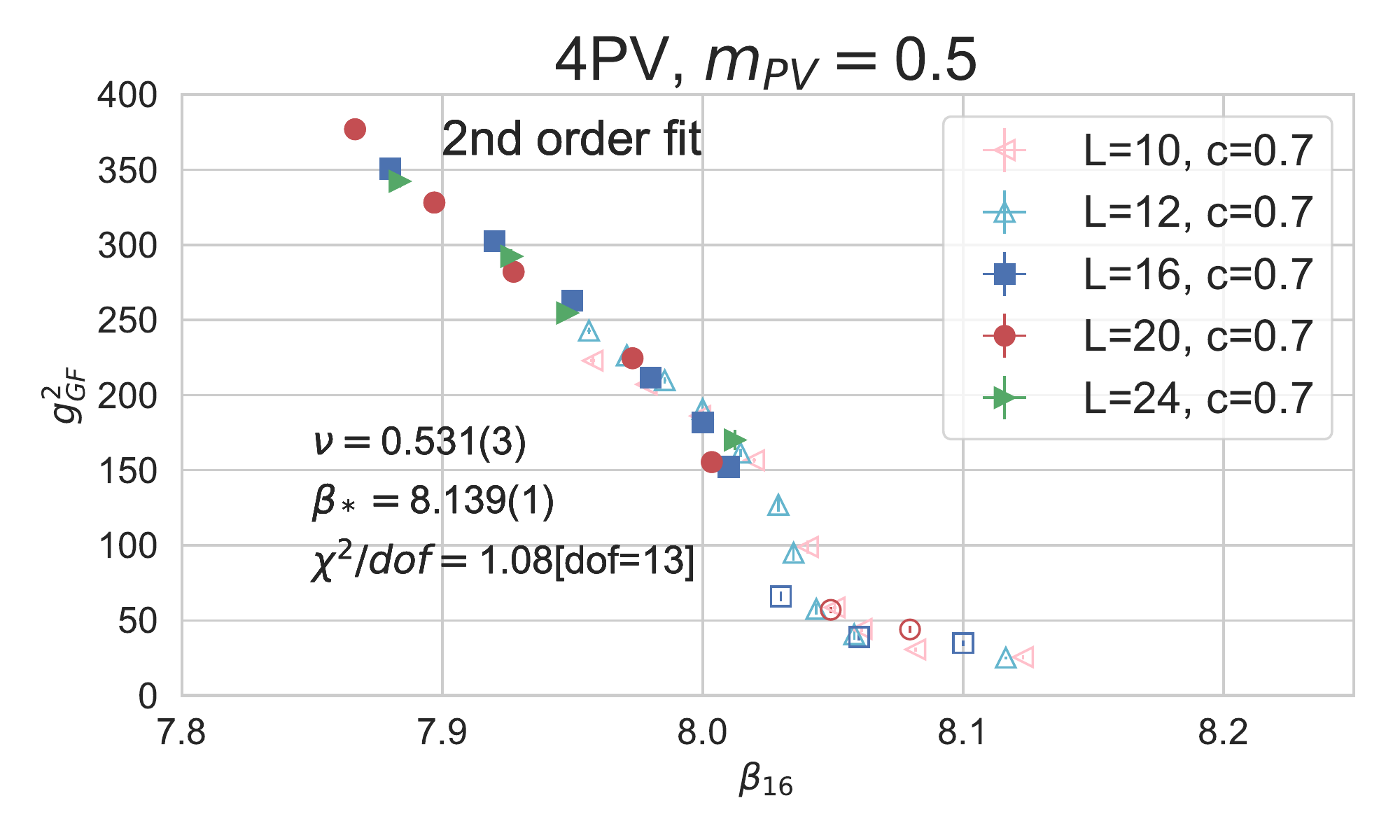} 
\vspace{-2ex}
\floatcaption{fig:2nd-order-cc}
{
Curve collapse analysis for  the various actions based on the second order scaling form of Eq. \ref{eq:scaling1}. The horizontal axis shows $\beta_{16}(\beta;L) = (L/L_0)^{1/\nu} (\beta_\star -\beta)$, $L_0=16$, i.e. the transformed coupling $\beta$  according  to Eq. \ref{eq:b0_2nd}. 
Only data points marked by filled symbols are included in the fit.}
 \end{figure*}
%%%%%%%%%%%%%%%%%%%%%%%%%%%%%%%%%%%%%%%%%%%%%%%%%%%%%%%%%%%%%%%%%%%%%%%%% 
%%%%%%%%%%%%%%%%%%%%%%%%%%%%%%%%%%%%%%%%%%%%%%%%%%%%%%%%%%%%%%%%%%%%%%%%%
\begin{figure*}[t]
%\vspace{-1ex}
%\hspace{-15mm}
\includegraphics[width=0.99\columnwidth]{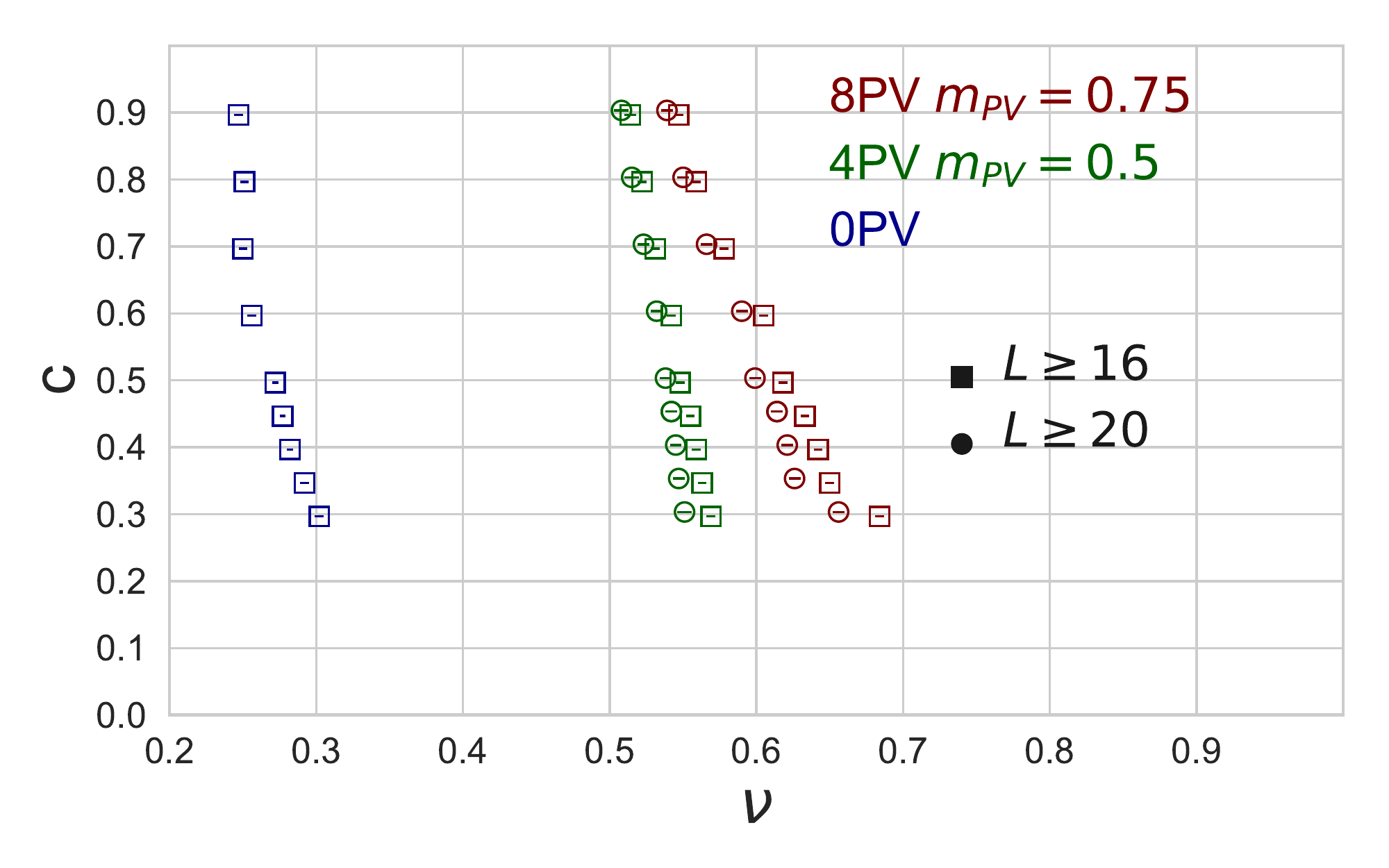} 
\includegraphics[width=0.99\columnwidth]{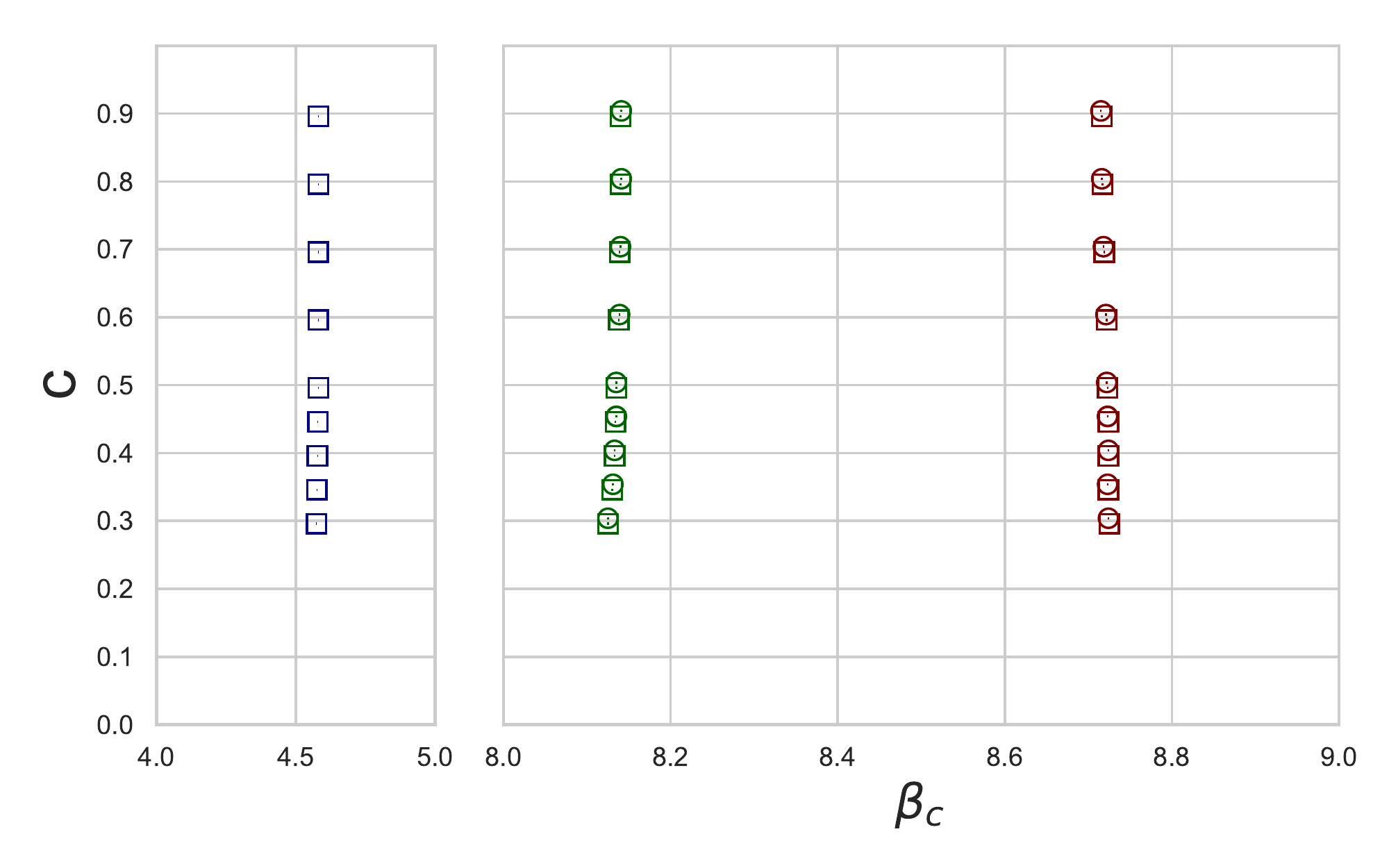} 
\vspace{-2ex}
\floatcaption{fig:2nd-order-summary}
{
Summary of the curve collapse analysis for  the various actions based on the second order scaling form of Eq. \ref{eq:scaling1}. The left panel shows the  the critical exponent $\nu$, the right panels  the critical coupling $\beta_\star$ as the parameter $c=\sqrt{8t}/L$ varies.  $\beta_\star$ is the infinite volume critical coupling and should be compared to $\beta$ in Figs. \ref{fig:0PV} - \ref{fig:4PV-m0.5}, not to $\beta_{16}$ in Fig. \ref{fig:2nd-order-cc}.    }
 \end{figure*}
%%%%%%%%%%%%%%%%%%%%%%%%%%%%%%%%%%%%%%%%%%%%%%%%%%%%%%%%%%%%%%%%%%%%%%%%% 

If the phase transition  is continuous, it should follow a scaling form like Eq. \ref{eq:scaling1} or Eq. \ref{eq:scaling2}.  A first order phase transition should follow the scaling of Eq. \ref{eq:scaling1} with exponent $\nu=1/4$. The critical coupling $\beta_\star$, exponent $\nu$ and parameter $\zeta$ can be determined based on a curve collapse analysis of $\gGF(x; c)$ vs. $x=|\beta_\star-\beta|^{\nu}\,L$ or $x= L \text{exp}\left(-\zeta |\beta/\beta_\star-1|^{-\nu}\right)$, respectively. I perform this analysis similar the Nelder-Mead method as described in \cite{curvecollape:2001}. 

Specifically, I chose a reference volume $L_0$ and interpolate $g_{GF}^2(\beta,L_0;c)$  versus $\beta$ with a (smoothed) cubic spline.  In the analysis below I use $L_0=16$. I fit the parameters $\beta_\star$, $\nu$ (and $\zeta$) by minimizing the deviation of $g_{GF}^2(\beta,L;c)$  from this interpolating curve at $\beta_{0}$, where $\beta_{0}$ is the solution of the equation
\be
L^{1/\nu} ( \beta_\star-\beta) = L_0^{1/\nu} ( \beta_\star-\beta_0) \, ,
\label{eq:b0_2nd}
\ee
if fitting Eq. \ref{eq:scaling1}  or
\be
 L \, \text{exp}(-\zeta |\beta/\beta_\star - 1|^{-\nu})  = L_0 \, \text{exp}(-\zeta |\beta_0/\beta_\star - 1|^{-\nu}) \, ,
\label{eq:b0_BKT}
\ee
if fitting Eq. \ref{eq:scaling2} The procedure can be iterated by refining the ``ideal" interpolating curve using more volumes, once the fit parameters are known approximately.

\subsubsection{Second order scaling test}\label{sect:2nd-order}

Fig. \ref{fig:2nd-order-cc} shows the curve collapse analysis for the three actions based on the second order scaling form of Eq. \ref{eq:scaling1} at $c=0.45$ and $c=0.70$.  I included only  $L\ge 16$ in the fit.  In addition, I excluded those data points that are not clearly in the S4 phase\footnote{
At a first order phase transition the critical coupling of an MCRG approach can overshoot the discontinuity, especially if the transition is strongly first order \cite{Decker:1987mu,Hasenfratz:1987mv}. In order to avoid  data that are in the weak coupling phase I employ a conservative cut based on the S4 order parameter.}. 
 However, for completeness,  I show the smaller volumes and excluded data  points in   Fig.  \ref{fig:2nd-order-cc}  using open symbols.  
 The top left panel shows the curve collapse for the 0PV action.  The fit with $c=0.45$ predicts  $\nu=0.285(2)$,  close to the expected first order 
 $\nu=0.25$ value with $\chi^2/\text{dof}=1.09$. The fit with $c=0.70$ predicts $\nu=0.255(3)$, even closer to the expected discontinuity exponent, with similarly small $\chi^2/\text{dof}$.   The predicted  critical couplings are consistent, 
 $\beta_\star=4.584(1)$ and 4.585(1), and compatible with the data in Fig. \ref{fig:0PV} and the observed phase transition in Ref.\cite{Schaich:2015psa}.
 The bottom two panels show the curve collapse for the two PV improved actions.  Again, only $L\ge16$ are included in the fit. In all cases $\chi^2/\text{dof}$ is $\mathcal{O}(1)$, but the predicted exponent $\nu$ varies between 0.53 and 0.63, very different from the discontinuity exponent of a first order phase transition. 
 
To test the consistency of the curve collapse, I repeat the analysis varying the parameter $c=\sqrt{8t}/L$ between 0.3 and 1.0. I also compare fits using  volumes $L\ge 16$ or  $L\ge 20$ for the PV actions. Fig. \ref{fig:2nd-order-summary} summarizes the predicted  $\nu$ and corresponding $\beta_\star$ values for the three actions.  The observable drift at smaller $c$ indicates that the RG flow has not yet reached the vicinity of the renormalized trajectory. The 0PV action stabilizes at $c > 0.5$ and predicts  $\nu=0.25(1)$, the expected scaling exponent at a first order transition.
 The exponent $\nu$ for the PV improved  actions  drift and show significant variation depending on the action even at $c\approx 0.9$.  I do not find a consistent scaling fit with volumes $L \ge 16$, though it is possible that on larger volumes the curve collapse analysis would predict a consistent exponent, possibly around $\nu=0.5$. 
 For both PV improved actions, a first order transition is  not consistent with the numerical data.

%%%%%%%%%%%%%%%%%%%%%%%%%%%%%%%%%%%%%%%%%%%%%%%%%%%%%%%%%%%%%%%%%%%%%%%%%
\begin{figure*}[t]
%\vspace{-1ex}
%\hspace{-15mm}
\includegraphics[width=0.99\columnwidth]{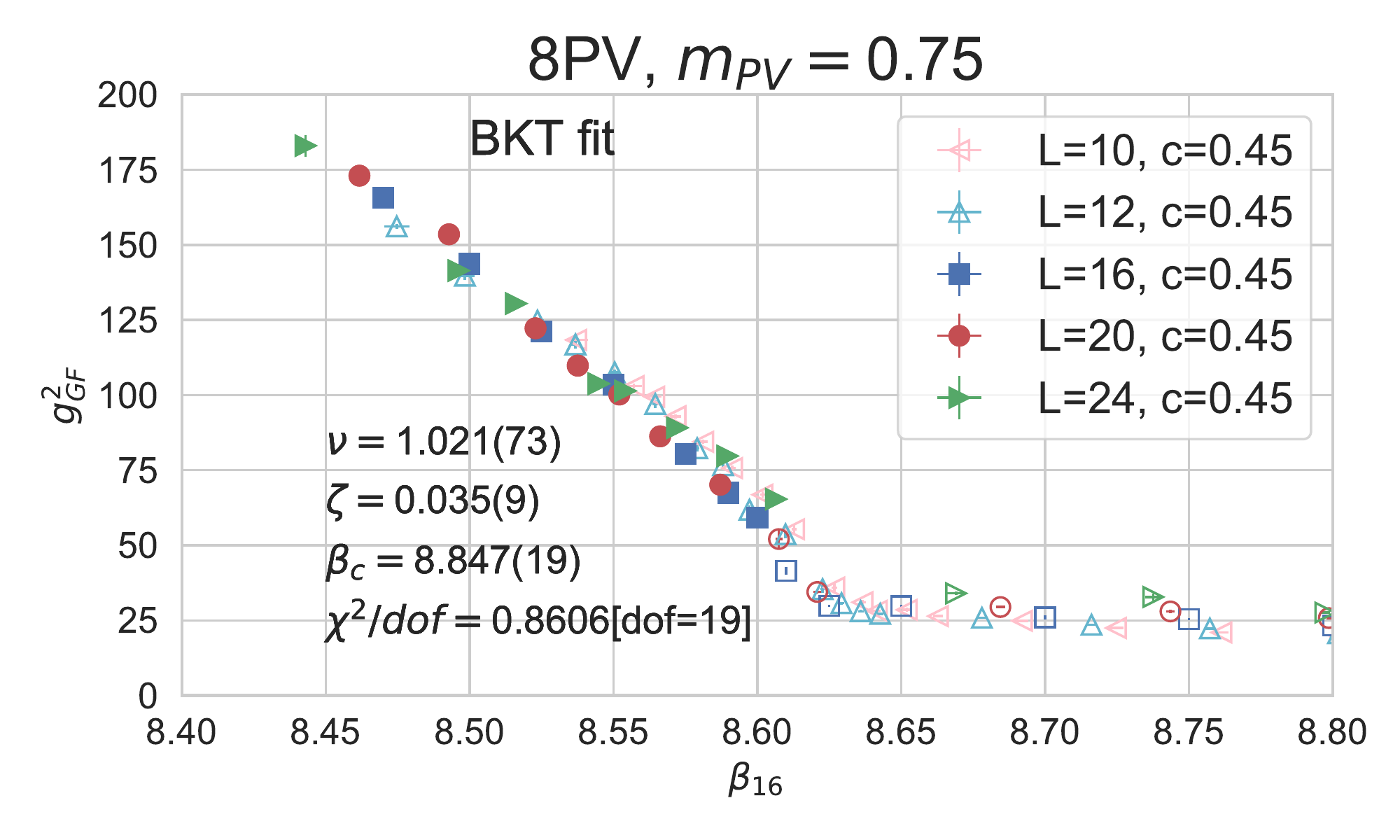} 
\includegraphics[width=0.99\columnwidth]{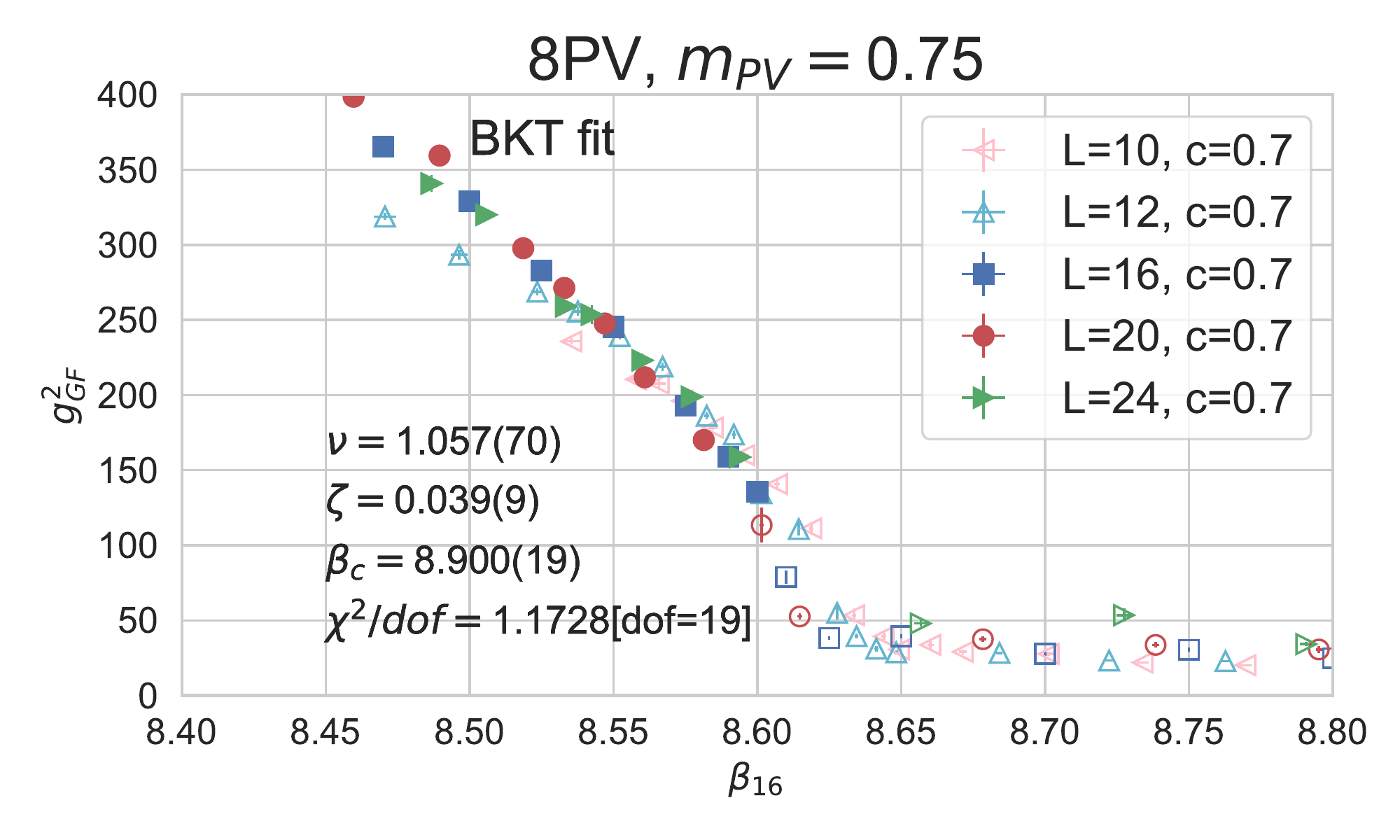} \\
\includegraphics[width=0.99\columnwidth]{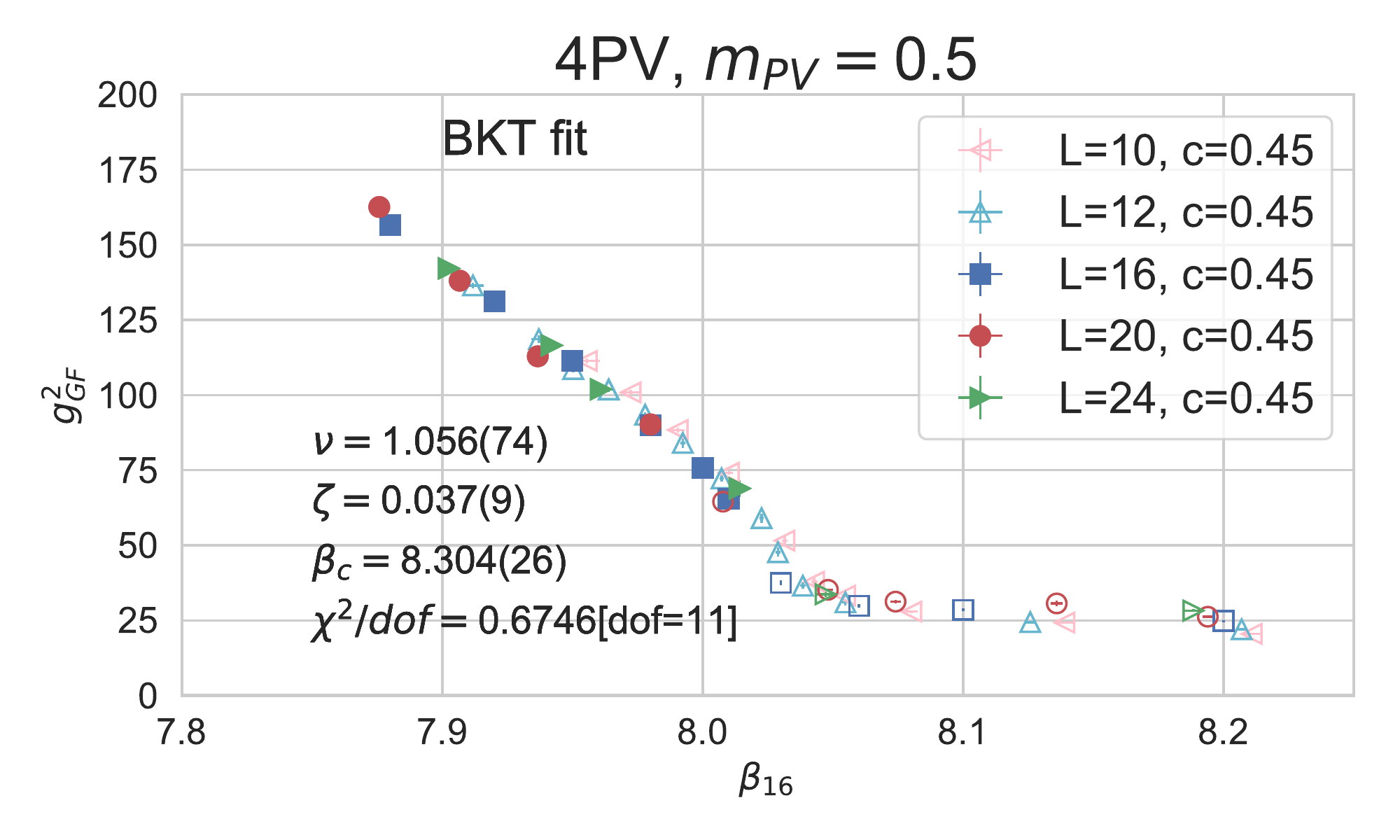} 
\includegraphics[width=0.99\columnwidth]{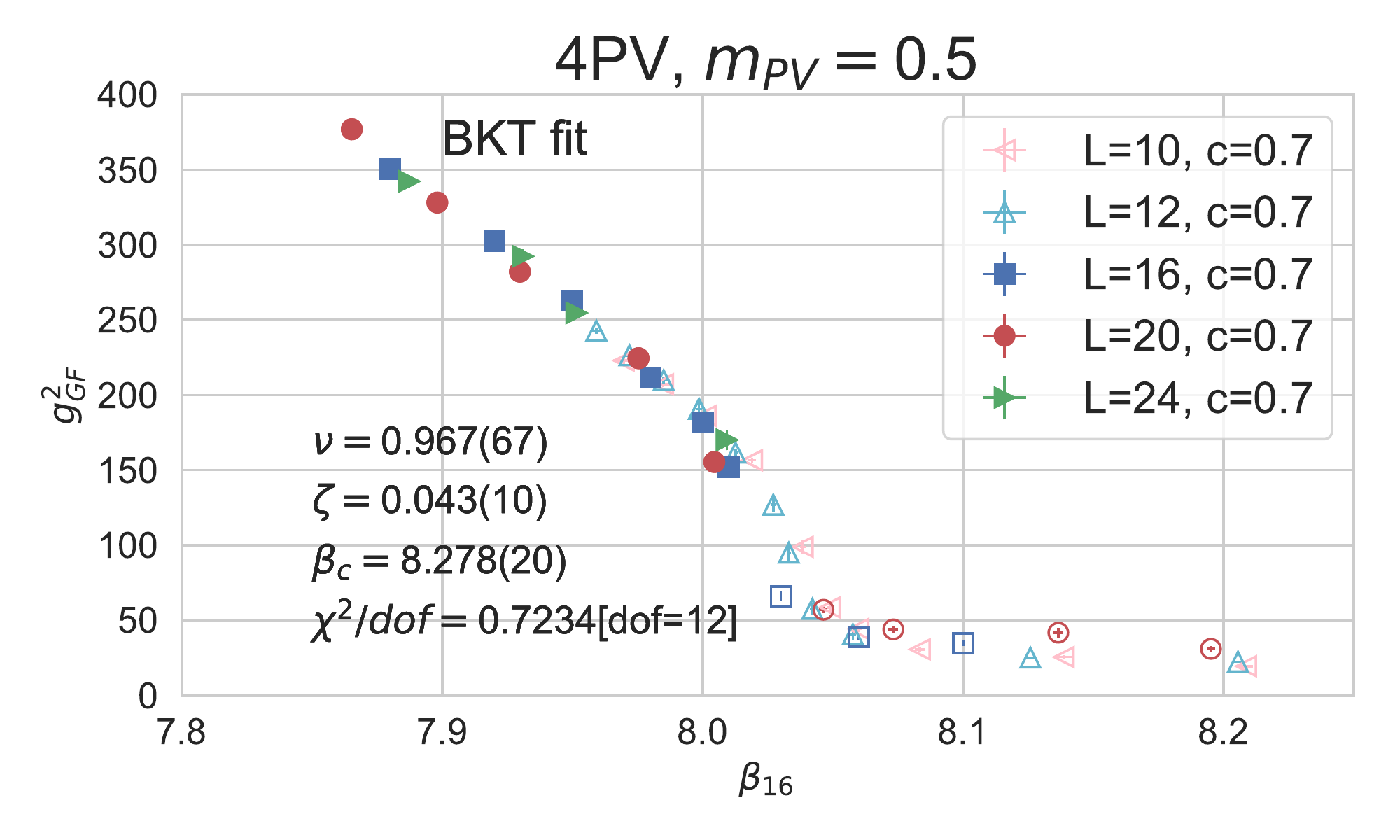} \\

\vspace{-2ex}
\floatcaption{fig:BKT-1}
{
Curve collapse analysis for  the PV  actions based on the BKT scaling form of Eq. \ref{eq:scaling2} with $c=0.45$ (left panels) and $c=0.70$ (right panels). Only  $L\ge 16$ volumes are included in the fit. Data from smaller volumes or outside the S4 phase are  shown with open symbols. $\beta_{16}(\beta;L)$ is the transformed coupling  according  to Eq. \ref{eq:b0_BKT}.
}
 \end{figure*}
%%%%%%%%%%%%%%%%%%%%%%%%%%%%%%%%%%%%%%%%%%%%%%%%%%%%%%%%%%%%%%%%%%%%%%%%% 

%%%%%%%%%%%%%%%%%%%%%%%%%%%%%%%%%%%%%%%%%%%%%%%%%%%%%%%%%%%%%%%%%%%%%%%%%
\begin{figure*}[t]
%\vspace{-1ex}
%\hspace{-15mm}
\includegraphics[width=0.5\columnwidth]{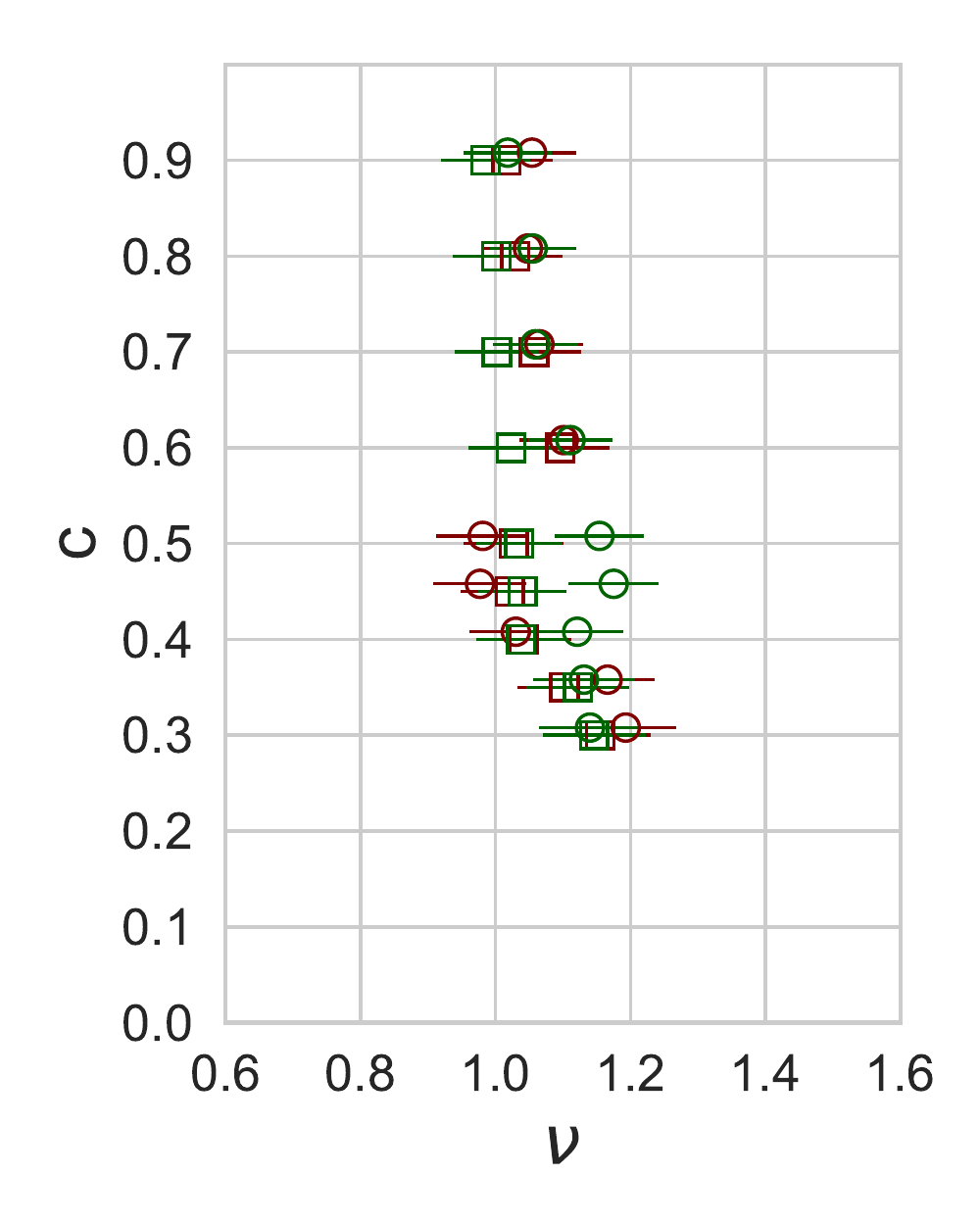} 
\includegraphics[width=0.375\columnwidth]{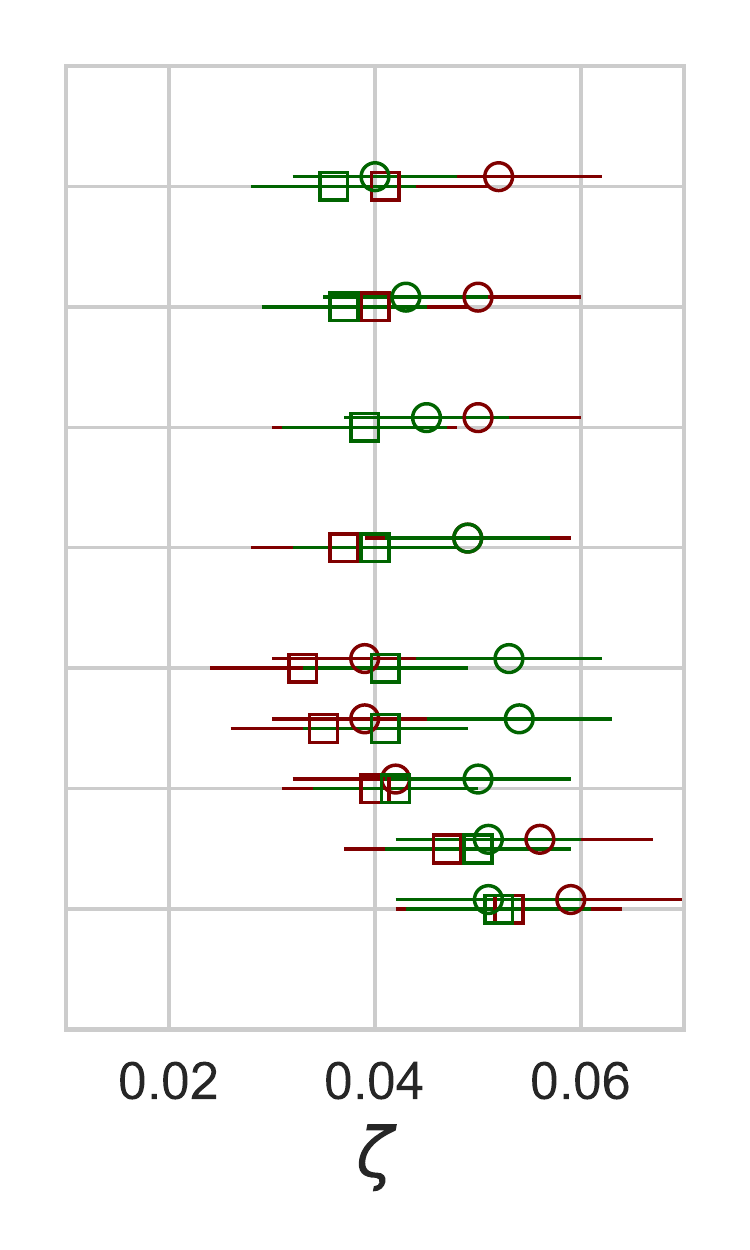} 
\includegraphics[width=0.75\columnwidth]{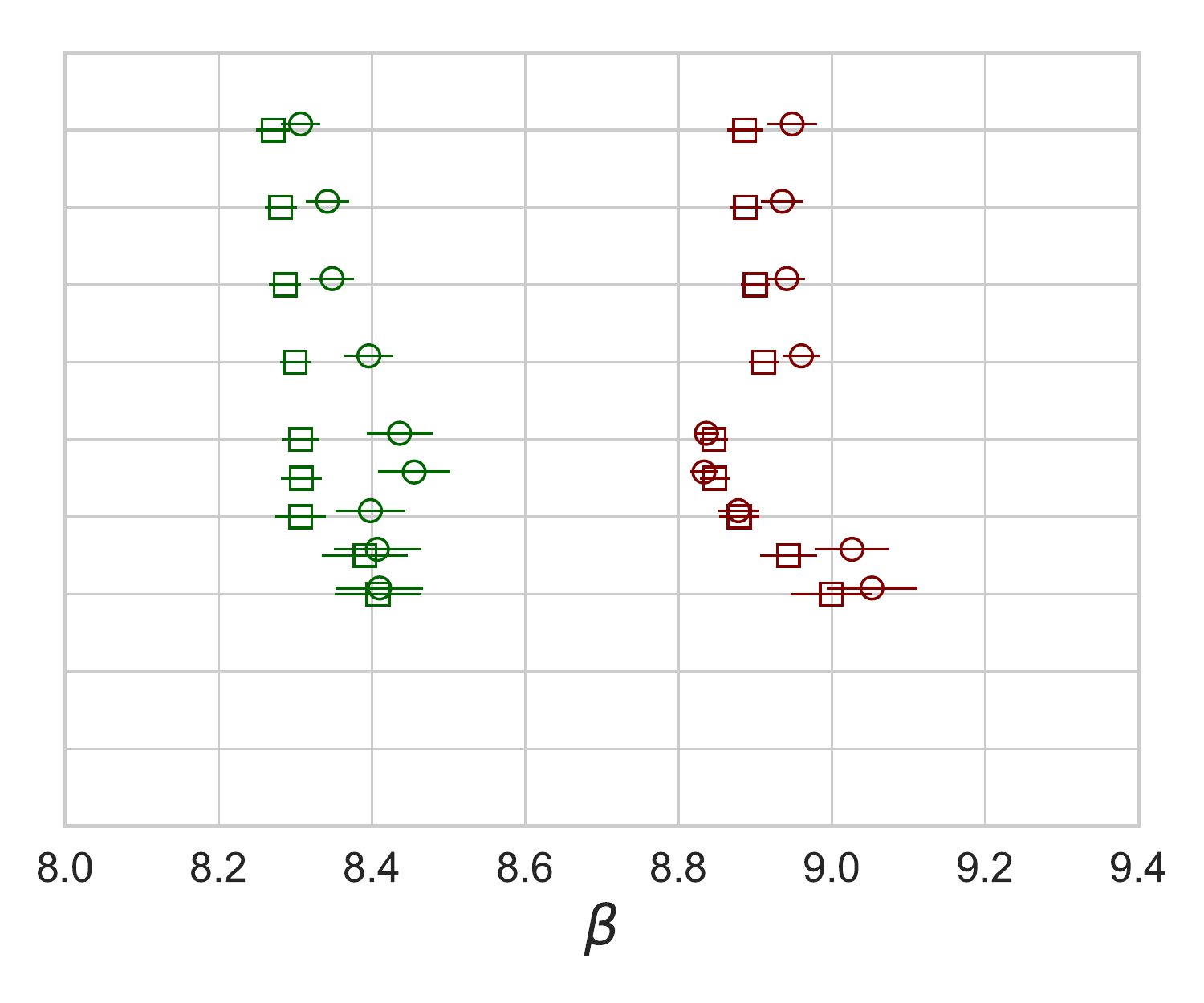} 
\includegraphics[width=0.375\columnwidth]{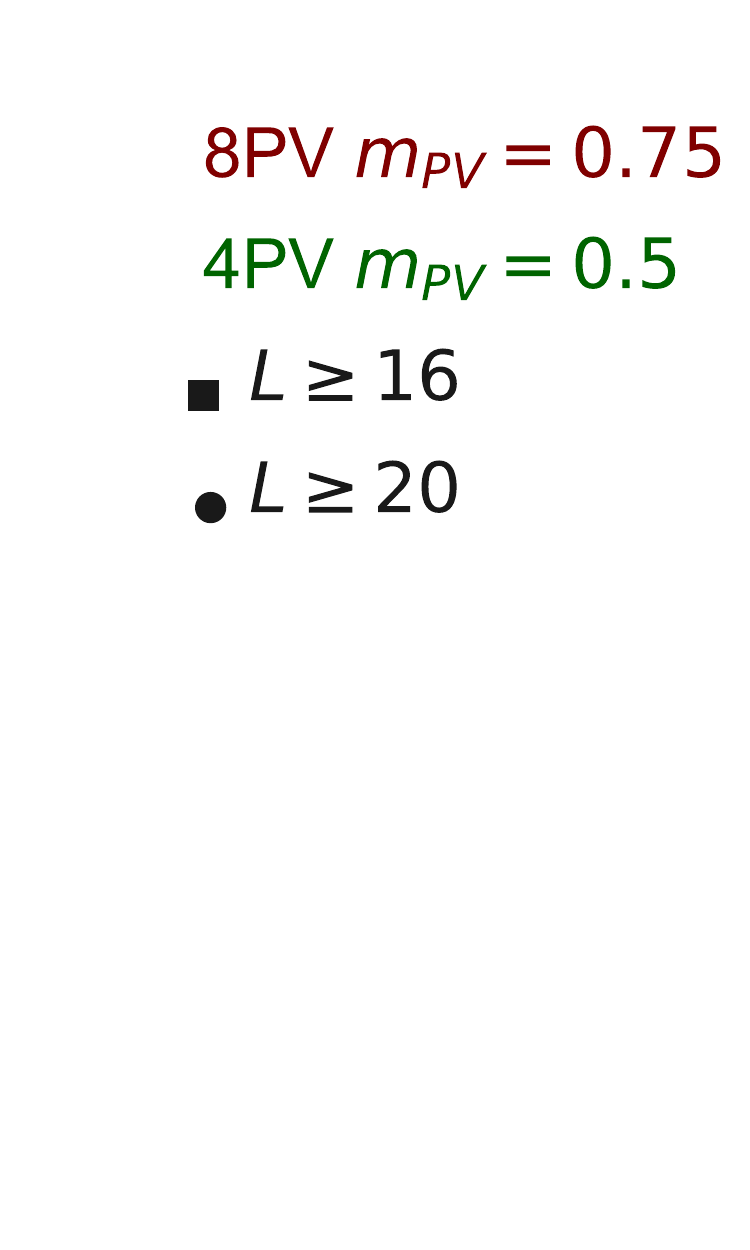} \\
\vspace{-2ex}
\floatcaption{fig:BKT-summary}
{
Summary of the curve collapse analysis for  the 8PV-m0.75 and 4PV-m0.5 actions based on the BKT scaling form of Eq. \ref{eq:scaling2}. 
}
 \end{figure*}
%%%%%%%%%%%%%%%%%%%%%%%%%%%%%%%%%%%%%%%%%%%%%%%%%%%%%%%%%%%%%%%%%%%%%%%%% 
%%%%%%%%%%%%%%%%%%%%%%%%%%%%%%%%%%%%%%%%%%%%%%%
\subsection{\label{fss} Finite size scaling analysis}
%%%%%%%%%%%%%%%%%%%%%%%%%%%%%%%%%%%%%%%%%%%%%%%

\subsubsection{``Walking"  scaling test}\label{sect:BKT}  

The ``walking" or BKT-like scaling form given in Eq. \ref{eq:scaling2}  assumes that the RG $\beta$ function just touches zero. This fit form has three free parameters, though $\nu=1.0$ is the theoretically motivated value  \cite{Kaplan:2009kr}. Using $L\ge16$ volumes I am able to fit all three parameters.   Fig. \ref{fig:BKT-1}  shows the result of the curve collapse fits for both  PV actions at $c=0.45$ and 0.70.  As before, the open symbol data points are not included in the fit. All fits predict $\nu$ consistent with the expected 1.0 value with small $\chi^2/\text{dof}$ and consistent $\zeta$ and $\beta_\star$ values.

The predicted  fit parameter values are shown in Fig. \ref{fig:BKT-summary} for different $c$ values in the range $c=0.3 \--1.0$ and volumes $L\ge 16$  and $L\ge 20$. They all show consistent results. 

Fits with  BKT-like scaling have smaller $\chi^2/\text{dof}$ than the second-order scaling. The predicted $\nu\approx 1.0$ exponent and no significant dependence on $c$ also favors the ``walking" scaling scenario.  However the second order scaling form discussed in Sect. \ref{sect:2nd-order} is not  excluded. It is well known that corrections to BKT scaling in the  2-dimensional XY model are  very large \cite{Hasenbusch:2005xm}.
In the 8-flavor system I have concentrated on fitting the leading terms only. More sophisticated curve collapse fitting, better statistics, and especially larger volumes might allow distinguishing the two cases in the future \cite{Sale:2021xsq}.

\section{\label{summary} Summary}
%%%%%%%%%%%%%%%%%%%%%%%%%%

I have investigated  the SU(3) gauge model with two sets of staggered fermions in the chiral limit.  Staggered lattice fermions are equivalent to  Dirac-K\"ahler fermions, and in the chiral limit two staggered fields correspond to 16 Weyl spinors. This system could lead to symmetric mass generation if appropriate  terms are induced by the strong interaction.

I study two improved lattice  actions that contain heavy Pauli-Villars type bosonic fields. Both improved actions shift the first order phase transition of the original  system to weaker bare couplings and smooth out the transition. Using finite size scaling methods I showed  that the phase transition with both PV improved actions appears continuous. I investigated scaling according to standard second order phase transition, and also ``walking" or BKT-like scaling.  First order phase transition with $\nu=1/4$ is not consistent with the numerical data.  While I cannot  exclude second order scaling, the curve collapse fit favors ``walking" scaling with $\nu\approx 1.0$.
A ``walking" scaling phase transition could  imply that the system is at the sill of the conformal window as sketched on the right panel of Fig. \ref{fig:phasediag}.
If that is indeed the case, it is likely that the complete t'Hooft anomaly cancellation of
the lattice formulation plays an essential role.

The weak coupling phase with all three actions appears  chirally symmetric in the  volumes considered, i.e. the simulations are well controlled even in the $am_f=0$ limit with meson spectrum showing chiral symmetry. Simulations at finite fermion mass that contrast the $\delta$-regime rotator spectrum with the mass-deformed hyperscaling prediction might be required to establish the infrared properties of the weak coupling phase~\cite{Hasenfratz:2009mp,Fodor:2009wk}.  In this work I concentrate on the order of the phase transition from the strong coupling side and do not directly investigate the infrared properties of the weak coupling phase.  The strong coupling regime breaks the single site shift symmetry of the staggered action. The S4 phase could describe symmetric mass generation, where confinement leads to mass generation without breaking chiral symmetry.  The existence of a continuous phase transition would allow taking an infinite cutoff continuum limit. This very exciting possibility requires further analytical and numerical studies.  

Lattice studies with $N_f=12$ fundamental flavors also show an S4 phase in the strong coupling. However, my preliminary numerical studies with various PV actions did not show continuous phase transition in that system. t'Hooft anomalies do not cancel with 12 flavors (i.e. three staggered fermions), possibly explaining the difference between 12 and 8 flavors. However, further studies are needed.

If the model with 2 staggered fermions is conformal in the weak coupling regime, 8 fundamental Dirac fermions should also be conformal. The critical properties at the continuous phase transition, however, are not necessarily universal. It would be very interesting to investigate 8 fundamental flavors with improved domain wall fermions that allow the study of the strong gauge couplings where the continuous phase transition with staggered fermions occur.

\vspace{2ex}
\section*{Acknowledgments}
%%%%%%%%%%%%%%%%%%%%%%%%%%
The phase diagram depicted in Fig. \ref{fig:phasediag} came from a discussion with S. Rychkov. I am indebted for his insight that lead to the present investigation.
 I am grateful  to Yigal Shamir for his constructive criticism and suggestions, and to Simon Catterall for discussions 
 about t'Hooft anomaly cancellation and SMG. 
 I have presented preliminary results to the LSD collaboration and benefited from their probing questions when finalizing the manuscript. I thank Oliver Witzel who carefully read an earlier version of this manuscript for his many comments and suggestions.

Computations for this work were carried out in part on facilities of the USQCD Collaboration, which are funded by the Office of Science of the U.S.~Department of Energy, the RMACC Summit supercomputer \cite{UCsummit}, which is supported by the National Science Foundation (awards No.~ACI-1532235 and No.~ACI-1532236), the University of Colorado Boulder, and Colorado State University,  and on  the University of Colorado Boulder HEP Beowulf cluster. I thank the University of Colorado for providing the facilities essential for the completion of this work.
My reseach is  supported by DOE grant DE-SC0010005.

%%%%%%%%%%%%%%%%%%%%%%%%%%%%%%%%%%%%%%%%%%%%%%%%%%%%%%%%%%%%%%%%%%%%%%%%%%
%%%  figures we didn't use
%%%%%%%%%%%%%%%%%%%%%%%%%%%%%%%%%%%%%%%%%%%%%%%%%%%%%%%%%%%%%%%%%%%%%%%%%%
%\bibliographystyle{apsrev4-2}
\bibliography{/Users/anna/Dropbox/Text/PV_bosons/PV}

\end{document}